\documentclass[]{elsarticle} %review=doublespace preprint=single 5p=2 column
\usepackage[hyphens]{url}

  \journal{eLife Sciences} % Sets Journal name

\providecommand{\tightlist}{%
  \setlength{\itemsep}{0pt}\setlength{\parskip}{0pt}}

\usepackage{graphicx}
\usepackage{booktabs} % book-quality tables
\usepackage[T1]{fontenc}
\usepackage{lmodern}
\usepackage{mathpazo}
\usepackage[onehalfspacing]{setspace}
\usepackage{amssymb,amsmath}
\usepackage{ifxetex,ifluatex}
\usepackage{fixltx2e} % provides \textsubscript
\IfFileExists{upquote.sty}{\usepackage{upquote}}{}
\ifnum 0\ifxetex 1\fi\ifluatex 1\fi=0 % if pdftex
  \usepackage[utf8]{inputenc}
\else % if luatex or xelatex
  \usepackage{fontspec}
  \ifxetex
    \usepackage{xltxtra,xunicode}
  \fi
  \defaultfontfeatures{Mapping=tex-text,Scale=MatchLowercase}
  
\fi
\IfFileExists{microtype.sty}{\usepackage{microtype}}{}
\ifxetex
  \usepackage[setpagesize=false, % page size defined by xetex
              unicode=false, % unicode breaks when used with xetex
              xetex]{hyperref}
\else
  \usepackage[unicode=true]{hyperref}
\fi
\hypersetup{breaklinks=true,
            bookmarks=true,
            pdfauthor={},
            pdftitle={Evaluating distributional regression strategies for modelling self-reported sexual age-mixing},
            colorlinks=false,
            urlcolor=blue,
            linkcolor=magenta,
            pdfborder={0 0 0}}
\newlength{\csllabelwidth}
\newlength{\cslhangindent}
  {}%
  {\par}
\newenvironment{CSLReferences}[3] % #1 hanging-ident, #2 entry spacing
 {% don't indent paragraphs
  \setlength{\parindent}{0pt}
  \ifodd #1 \everypar{\setlength{\hangindent}{\cslhangindent}}\ignorespaces\fi
  \ifnum #2 > 0
  \setlength{\parskip}{#2\baselineskip}
  \fi
 }%
 {}
\usepackage{calc} % for calculating minipage widths

\usepackage{float}
\usepackage{booktabs}
\usepackage{longtable}
\usepackage{array}
\usepackage{multirow}
\usepackage{wrapfig}
\usepackage{colortbl}
\usepackage{pdflscape}
\usepackage{tabu}
\usepackage{threeparttable}
\usepackage{threeparttablex}
\usepackage[normalem]{ulem}
\usepackage{makecell}

\begin{document}
\begin{frontmatter}

  \title{Evaluating distributional regression strategies for modelling
self-reported sexual age-mixing}
    \author[Maths]{Timothy M Wolock\corref{1}}
   \ead{t.wolock18@imperial.ac.uk} 
    \author[Maths]{Seth R Flaxman}
  
    \author[DIDE,LSHTM]{Kathryn A Risher}
  
    \author[Manicaland]{Tawanda Dadirai}
  
    \author[DIDE,Manicaland]{Simon Gregson}
  
    \author[DIDE]{Jeffrey W Eaton}
  
      \address[Maths]{Department of Mathematics, Imperial College
London, London, UK}
    \address[DIDE]{MRC Centre for Global Infectious Disease Analysis,
School of Public Health, Imperial College London, London, UK}
    \address[LSHTM]{London School of Hygiene \& Tropical Medicine,
London, UK}
    \address[Manicaland]{Manicaland Centre for Public Health Research,
Biomedical Research and Training Institute, Harare, Zimbabwe}
      \cortext[1]{Corresponding Author}
  
  \begin{abstract}
  The age dynamics of sexual partnership formation determine patterns of
  sexually transmitted disease transmission and have long been a focus
  of researchers studying human immunodeficiency virus. Data on
  self-reported sexual partner age distributions are available from a
  variety of sources. We sought to explore statistical models that
  accurately predict the distribution of sexual partner ages over age
  and sex. We identified which probability distributions and outcome
  specifications best captured variation in partner age and quantified
  the benefits of modelling these data using distributional regression.
  We found that distributional regression with a sinh-arcsinh
  distribution replicated observed partner age distributions most
  accurately across three geographically diverse data sets. This
  framework can be extended with well-known hierarchical modelling tools
  and can help improve estimates of sexual age-mixing dynamics.
  \end{abstract}
  
 \end{frontmatter}

\hypertarget{introduction}{%
\section{Introduction}\label{introduction}}

Patterns in sexual mixing across ages determine patterns of transmission
of sexually transmitted infections (STIs). Consequently, sexual
age-mixing has been of great interest to researchers studying the human
immunodeficiency virus (HIV) since the beginning of the global epidemic.
Anderson et al. (1992) used a model of partnership formation to predict
that mixing between young women and older men would amplify the
already-substantial effect on HIV on population growth. Garnett \&
Anderson (1994) used a mathematical model to show that patterns of
age-mixing could substantially influence the magnitude and timing of
hypothetical epidemic trajectories, while Hallett et al. (2007)
demonstrated that delaying sexual debut and increasing age-similar
partnerships could reduce an individual's risk of HIV infection in a
highly endemic setting.

These modelling studies have been complemented by analyses of survey and
population cohort data on age-mixing patterns. Gregson et al. (2002)
observed that individuals with older partners were at greater risk of
HIV infection. Ritchwood et al. (2016) and Maughan-Brown et al. (2016)
found that larger age differences were associated with more risky sexual
behaviour in surveys of young South African people. On the other hand,
Harling et al. (2014) found that age-disparate relationships were not
associated greater risk of HIV acquisition in young women in South
Africa.

These results underscore the importance of considering age-mixing
dynamics when designing and evaluating HIV prevention strategies, and,
consequently, the importance of measuring them accurately. For example,
an intervention aiming to prevent new HIV infections among young women
could be undermined by high prevalence among older men. Identifying
changes in sexual partner age distributions and attributing them to
interventions might even be a valuable end by itself, in which case
accurate measurement must be complemented by an effective modelling
strategy.

Data about sexual partner age-mixing are routinely collected by
long-term cohort studies (such as those that comprise the ALPHA Network)
and large-scale household surveys (such as the Demographic and Health
Surveys) (Reniers et al., 2016; \emph{The {DHS} Program}, 2021).
Typically, these data consist of the respondent's age and sex and the
ages of their sexual partners in the last 12 months. These data are
highly variable, skewed, and often deviate substantially from
conventional parametric distributions, such as the normal distribution
or the gamma distribution (Beauclair et al., 2018).

One may consider statistical modelling approaches for the distribution
of partner age as a function of respondent age and sex. Some notable
previous approaches to modelling partner age distributions include
Hallett et al., who used a log-logistic distribution to model partner
age differences for women aged 15 to 45 years, assuming that the partner
age difference distributions did not vary over respondent age. More
recently, as an input to a model of \emph{Chlamydia trachomatis}, Smid
et al. (2018) fit skew normal distributions to each age-/sex-specific
partner age distribution and used a secondary regression model to smooth
the estimated skew normal parameters across respondent age. They
observed substantial changes in the estimated skew normal parameters
with respect to respondent age. Although this method allows for
non-linear variation across respondent age, their two-stage estimation
process makes uncertainty propagation complex. Replacing this process
with a single ``distributional'' regression model, in which all
distributional parameters (e.g.~the location, scale, skewness, etc.) are
modelled as functions of data (Kneib \& Umlauf, 2017), allows for
complex variation across respondent age while still robustly
incorporating uncertainty.

More broadly, no previous work has systematically evaluated the wide
variety of distributions potentially available to model partner age
distributions. These data are skewed, heavy-tailed, and
otherwise dissimilar to conventional statistical distributions due to
personal preferences, social dynamics, demographic change, and any
number of other factors. We were specifically interested in
distributions that introduce parameters to control tail weight, which
may capture intergenerational mixing that could sustain endemic HIV and
STI transmission. This led us to test the ability of the four-parameter
``sinh-arcsinh'' distribution originally proposed by Jones \& Pewsey
(2009) to fit to these data.

We hypothesized that integrating the sinh-arcsinh distribution into a
distributional modelling framework would allow us to replicate observed
partner age distributions more accurately than prior modelling
strategies. We tested this theory by comparing a variety candidate
strategies, which varied along three dimensions: the parametrisation of
the dependent variable, the choice of distribution, and the method for
incorporating variability across respondent age and sex.

\hypertarget{methods}{%
\section{Methods}\label{methods}}

We conducted two model comparison experiments to identify which of a set
of strategies best replicated partner age distributions. First, in our
probability distribution comparison, we identified which of a set of
distribution-dependent variable combinations fit best to
age-/sex-specific data subsets, and then, in our distributional
regression evaluation, we tested whether distributional regression
methods could be used to estimate age-/sex-specific partner age
distributions by sharing strength across observations. We divided the
model comparison into two separate experiments to make the probability
distribution comparison as fair as possible (accounting for the
possibility that certain distributions would perform particularly well
under certain regression specification).

\hypertarget{data}{%
\subsection{Data}\label{data}}

We analysed data on sexual partner age distributions from three sources:
the Africa Centre Demographic Information System, a health and
demographic surveillance site in uMkhanyakude district, South Africa
collected by the African Health Research Institute (AHRI) (Gareta et
al., 2021; Gareta et al., 2020a, 2020b), the Manicaland General
Population Cohort in Zimbabwe (Gregson et al., 2017), and the 2016-2017
Demographic and Health Survey (DHS) in Haiti (Institut Haïtien de
l'Enfance - IHE/Haiti \& ICF, 2018).

The AHRI and Manicaland studies are multi-round, open, general population
cohort studies designed to measure the dynamics of HIV, sexual risk
behaviour, and demographic change in sub-Saharan African settings. We
used rounds 1 through 6 of the Manicaland study, collected between 1998
and 2013. The AHRI data we used were collected annually between 2004 and
2018. The 2016-17 Haiti DHS was a large, nationally representative
household survey conducted in 2016 and 2017. We did not incorporate the
weights associated with the survey into this analysis because our primary
interest was in statistical modelling of partner age distribution as a
function of respondent age, not producing population representative
statistics for the Haitian population.

These data sets consisted of individuals' reports of their own age and
sex and the ages of each of their sexual partners from the last year.
Let \(i \in (1, .. ., N)\) index reported partnerships,
\(a_i \in [15, 64]\) and \(s_i\in\{0,1\}\) be the age and sex of the
respondent in partnership \(i\) with \(s=1\) indicating female, and
\(p_i\) be the age of non-respondent partner in partnership \(i\). These
questionnaires do not ask specifically about partner sex, but
self-reporting of non-heterosexual partnerships in these populations is
thought to be low.

Respondents in each of these data sets are disproportionately likely to
report that their partners' ages are multiples of five or multiples of
five away from their own age, leading to distinct ``heaping'' in the
empirical partner age (or age difference) distributions at multiples of
five. We tested the sensitivity of our results to heaping by developing
a simple ``deheaping'' algorithm, applying it to the AHRI data, and
running each analysis on the deheaped AHRI data. We present these
results in Appendix section ``Age heaping.''

\hypertarget{probability-distribution-comparison}{%
\subsection{Probability distribution
comparison}\label{probability-distribution-comparison}}

To identify the best probability distribution for modelling sexual
partner age distributions, we split each data set into 12 subsets by sex
and five-year age bin ranging from 20 to 50, resulting in 36 subsets,
and fit a number of distribution-dependent variable combinations to each
subset.

\hypertarget{distributions}{%
\paragraph{Distributions}\label{distributions}}

We tested five candidate probability distributions: normal, skew normal,
beta, gamma, and sinh-arcsinh. Table \ref{tab:densSummary} summarises
the domains, parameters, and probability density functions (PDFs) of
these distributions. Because the gamma distribution is always
right-skewed and men typically partner with women who are younger than
them, we transformed data among male respondents to be right-skewed when
using the gamma distribution. Specifically, we multiplied the men's
partners' ages by -1 to reflect the distribution horizontally across the
y-axis, and added 150 to the reflected ages to ensure that all resulting
values were positive. Similarly, the beta distribution is only defined
on the interval \((0,1)\), so, only when using a beta distribution, we
scaled all partner ages to be between zero and one using upper and lower
bounds of 0 and 150.

\begin{table}

\caption{\label{tab:densSummary}Details of the five distributions tested in this analysis. We define $x_z=(x-\mu)/\sigma$, $p(x)$ to be the standard normal PDF, $\Phi(x)$ to be the standard normal cumulative density function, $S_{\epsilon,\delta}(x) = \sinh(\epsilon + \delta\mathop{\mathrm{asinh}}(x))$, and $C_{\epsilon,\delta}=\cosh(\epsilon + \delta\mathop{\mathrm{asinh}}(x))$.}
\centering
\begin{tabular}[t]{>{\raggedright\arraybackslash}p{5em}l>{\raggedright\arraybackslash}p{3em}l}
\toprule
Distribution & Parameters & Domain & PDF\\
\midrule
Normal & $\begin{array}{l}
\mu \;\text{(location)}\\
\sigma>0 \;\text{(scale)}
\end{array}$ & $\mathbb{R}$ & $\frac{1}{\sigma\sqrt{2\pi}}\exp\left[\frac{-x_z}{2}\right]$\\
\hline
Skew normal & $\begin{array}{l}
\mu\;\text{(location)}\\
\sigma>0 \vphantom{1}\;\text{(scale)}\\
\epsilon\;\text{(skewness)}
\end{array}$ & $\mathbb{R}$ & $\frac{2}{\sigma}p(x_z)\Phi(\epsilon x_z)$\\
\hline
Gamma & $\begin{array}{l}
k >0\;\text{(shape)}\\
\theta>0\;\text{(scale)}
\end{array}$ & $\mathbb{R}^+$ & $\frac{1}{\Gamma(k)\theta^k}x^{k-1}\exp\left[\frac{-x}{\theta}\right]$\\
\hline
Beta & $\begin{array}{l}
\alpha >0\;\text{(left)}\\
\beta>0\;\text{(right)}
\end{array}$ & $\mathbb{R}^{(0,1)}$ & $\frac{x^{\alpha-1}(1-x)^{\beta-1}}{B(\alpha,\beta)}$\\
\hline
Sinh-arcinh & $\begin{array}{l}
\mu\;\text{(location)}\\
\sigma>0 \;\text{(scale)}\\
\epsilon \;\text{(skewness)}\\
\delta>0\;\text{(tail weight)}
\end{array}$ & $\mathbb{R}$ & $\frac{1}{\sigma\sqrt{2\pi}}\frac{\delta C_{\epsilon, \delta}(x_z)}{\sqrt{1+x_z^2}} \exp\left[-\frac{S_{\epsilon,\delta}(x_z)^2}{2}\right]$\\
\bottomrule
\end{tabular}
\end{table}

\begin{figure}
\includegraphics[width=1\linewidth]{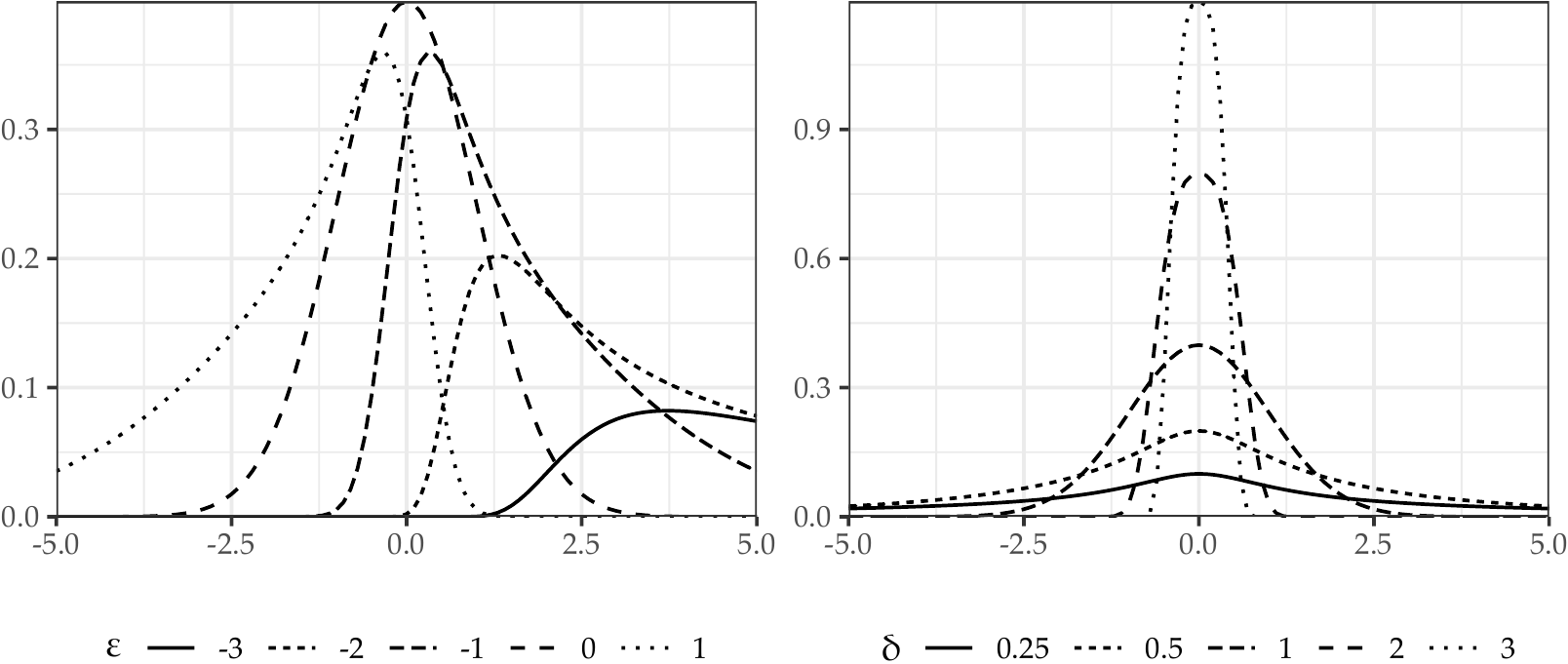} \caption{The sinh-arcsinh density with $\mu = 0$, $\sigma=1$, and a variety of assumptions about $\epsilon$ and $\delta$}\label{fig:sinhDens}
\end{figure}

The sinh-arcsinh distribution, presented by Jones \& Pewsey (2009), is
an extension of Johnson's \(S_U\) distribution (Johnson, 1949). It has
four parameters: location, scale, skewness, and tail weight (denoted,
\(\mu\), \(\sigma\), \(\epsilon\), and \(\delta\) respectively), and it
can deviate substantially from the normal distribution. Figure
\ref{fig:sinhDens} plots the density of this distribution with
\(\mu = 0\) and \(\sigma = 1\) for a variety of values of skewness and
tail weight.

\hypertarget{dependent-variable-transformations}{%
\paragraph{Dependent variable
transformations}\label{dependent-variable-transformations}}

We considered the possibility that certain distributions could interact
well with particular transformations of the dependent variable (partner
age) by testing a set of four potential outcome parametrisations. For
example, if \(X\) is a positive-valued, right-skewed random variable,
then assuming \(\log X\) is normally distributed might be more effective
than assuming that \(X\) itself is normal.

Let \(y_i\) be the dependent variable value for partnership \(i\), and
let \(a_i\) and \(p_i\) be the respondent age and partner age of
partnership \(i\), respectively. We tested the following dependent
variables:

\begin{enumerate}
\def\labelenumi{\arabic{enumi}.}
\tightlist
\item
  \textbf{Linear age}: \(y_i=p_i\). This is untransformed partner age,
  included as a baseline. It has the undesirable quality of being able
  to predict negative ages.
\item
  \textbf{Age difference}: \(y_i = p_i-a_i\). If changes in expected
  partner age are consistent across respondent age then this variable
  would be more consistent across respondent age than the linear age.
  This parametrisation also allows for negative partner age predictions.
\item
  \textbf{Log-age}: \(y_i = \log p_i\). We can use a \(\log\) link
  function to ensure that our predictions will be positive-valued.
\item
  \textbf{Log-ratio}: \(y_i = \log(p_i/a_i)\). Finally, we can combine
  the link function and differencing approaches by modelling the log of
  the ratio of partner to respondent age. This variable will only
  produce positive predictions and should exhibit the same relative
  stability as the age difference variable.
\end{enumerate}

Because the gamma and beta distributions are not defined on the entire
real line, we only fit them with the linear age dependent variable with
the previously discussed transformations.

To identify which distribution-dependent variable combination best
modelled the characteristics of sexual partner age distributions, we
stratified each of our three data sets by sex and five-year age bin from
20-24 through 45-49. We fit every viable distribution-dependent variable
combination to all 36 subsets independently. Given that we fit only the
linear age dependent variable to the gamma and beta distributions,
comprising a total of 504 models (14 per data set). We fit each model
using the \texttt{brms} R package (Bürkner, 2018), defining custom
families as necessary.

\hypertarget{distributional-regression-evaluation}{%
\subsection{Distributional regression
evaluation}\label{distributional-regression-evaluation}}

Given a probability distribution that accurately replicated the
non-Gaussian characteristics of partner age distributions, we tested
whether or not distributional regression would allow us to pool data
across age and sex without sacrificing fit. In distributional regression, we
make all of our distributional parameters, not just the mean, functions
of data (Kneib \& Umlauf, 2017). Taking conventional Bayesian regression
as an example, we have

\[
\begin{array}{rcl}
y_i &\sim& \text{N}(\mu_i, \sigma) \\
\mu_i &=& \beta\mathbf{X}_i,
\end{array}
\]

where \(\beta\) and \(\log \sigma\) are free parameters. There is an
explicit assumption in this model that the standard deviation of the
generating distribution is constant across all observations. We can use
distributional regression to relax this assumption, making \(\sigma\) a
function of data:

\[
\begin{array}{rcl}
y_i &\sim& \text{N}(\mu_i, \sigma_i) \\
\mu_i &=& \beta^\mu\mathbf{X}^\mu_i \\
\log\sigma_i &=& \beta^\sigma\mathbf{X}^\sigma_i,
\end{array}
\]

where \(\beta^\mu\) and \(\beta^\sigma\) are now our free parameters.
Note that we have not assumed that \(\mathbf{X}^\mu=\mathbf{X}^\sigma\).
If \(\mathbf{X}^\sigma\) is a column of ones, this model is identical to
the conventional case. This approach increases the complexity of the
model and requires more data, but, based on previously described
characteristics of how the distribution of partnership age distribution
changes with age, even a simple model for our distributional parameters
could yield large improvements.

In this case, we used a sinh-arcsinh distribution and specified a model
for each of its four parameters. We fit a series of increasingly complex
distributional regression specifications to the three data sets using
\texttt{brms} (Bürkner, 2018), which has deep support for distributional
regression.

\begin{enumerate}
\def\labelenumi{\arabic{enumi}.}
\tightlist
\item
  \textbf{Conventional}: linear age-sex interaction for location and
  constants for all three higher-order parameters
\item
  \textbf{Distributional 1}: linear age-sex interaction for location and
  independent age and sex effects for all other parameters
\item
  \textbf{Distributional 2}: linear age-sex interactions for all four
  parameters
\item
  \textbf{Distributional 3}: sex-specific spline with respect to age for
  location and linear age-sex interactions for all other parameters
\item
  \textbf{Distributional 4}: sex-specific splines with respect to age
  for all four parameters
\end{enumerate}

Table \ref{tab:modelTab} describes all five models. By fitting a wide
set of specifications, we hoped to assess whether the additional
complexity incurred by distributional regression was valuable. More
detailed descriptions of each model are available in the ``Model
specification details'' section of the Appendix.

\begin{table}

\caption{\label{tab:modelTab}Summary of five models fit in this analysis.}
\centering
\begin{tabular}[t]{llll}
\toprule
Model & Distributional? & Location & Other parameters\\
\midrule
Conventional & No & Age-sex interaction & Constant\\
Distributional 1 & Yes & Age-sex interaction & Age and sex effects\\
Distributional 2 & Yes & Age-sex interaction & Age-sex interaction\\
Distributional 3 & Yes & Sex-specific splines & Age-sex interaction\\
Distributional 4 & Yes & Sex-specific splines & Sex-specific splines\\
\bottomrule
\end{tabular}
\end{table}

\hypertarget{model-comparison}{%
\subsection{Model comparison}\label{model-comparison}}

Across both analyses, we used two metrics to measure model fit. First,
we calculated the expected log posterior density (ELPD), which estimates
the density of the model at a new, unobserved data point (Vehtari et
al., 2017). In cases where we wanted to compare across dependent
variables, we multiplied the posterior densities of any variables
resulting from non-linear transformations of observed partner ages by
the Jacobians of the transformations. For example, if our observation
model was defined on the log-age dependent variable \(y_i = \log p_i\),
we divided the posterior density by \(p_i\). We used the \texttt{loo} R
package (Vehtari et al., 2020) to calculate ELPD values.

To measure the ability of our models to replicate partner age
distributions in an objective and interpretable way, we found the root
mean squared error (RMSE) between the observed and posterior predictive
quantiles. We calculated quantiles from 10 to 90 in increments of 10 by
age bin and sex in the data and in the posterior predictions, and found
the error in model prediction of each quantile. This measure tells how
well our model predicts the entire distribution in the same units as our
predictions. It is equivalent to finding the mean squared or median
absolute distance from the line of equality in a quantile-quantile (QQ)
plot.

\hypertarget{software}{%
\subsection{Software}\label{software}}

We conducted all of these analysis using the R programming language (R
Core Team, 2020) and the \texttt{brms} library (Bürkner, 2018). We used
the \texttt{loo} library to estimate all ELPDs (Vehtari et al., 2020),
and produced all plots in this paper with the \texttt{ggplot2} library
(Wickham, 2016). We cannot provide the data we used for this analysis, but we do provide code and data for an simulated case on GitHub (\url{https://github.com/twolock/distreg-illustration}).

\hypertarget{results}{%
\section{Results}\label{results}}

\begin{figure}
\includegraphics[width=1\linewidth]{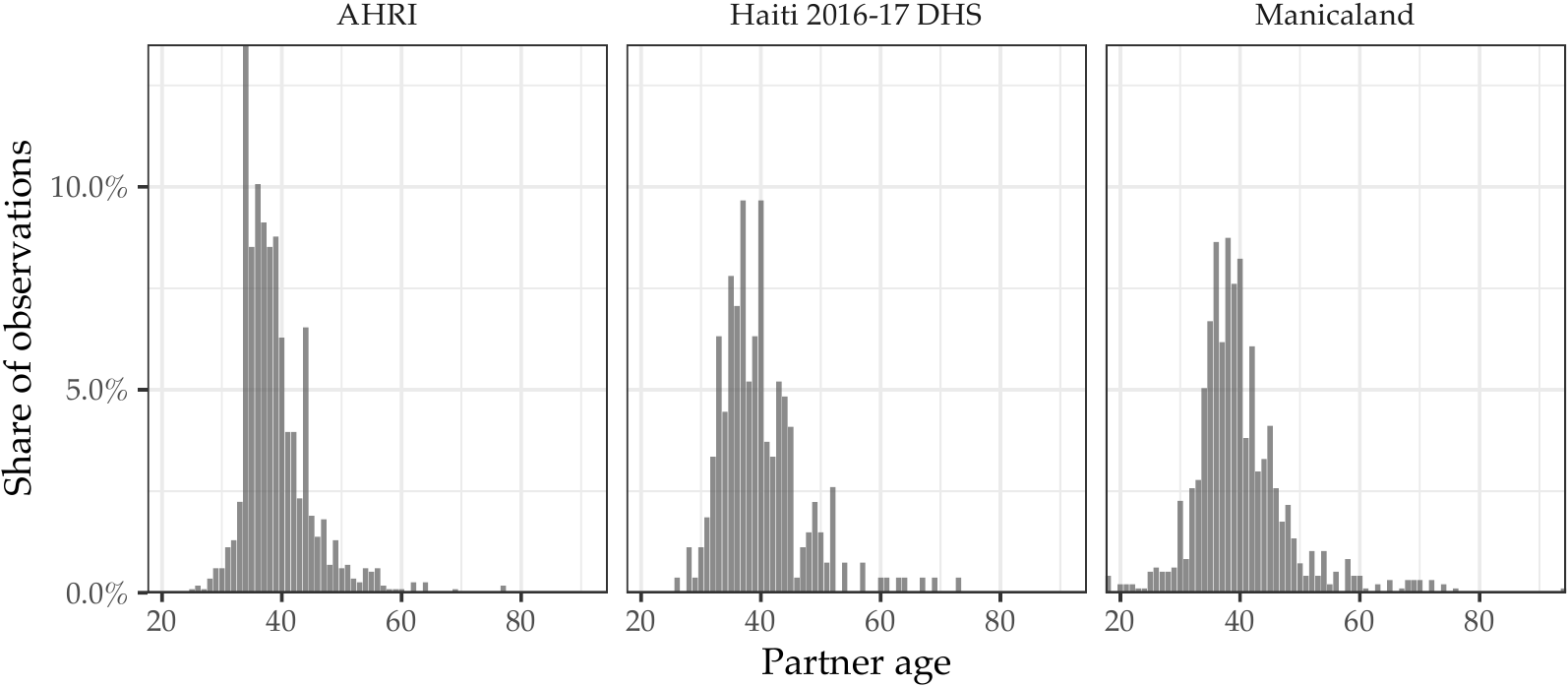} \caption{Observed partner age distributions among women aged 34 years in all three data sets.}\label{fig:studyHist}
\end{figure}

\begin{figure}
\includegraphics[width=1\linewidth]{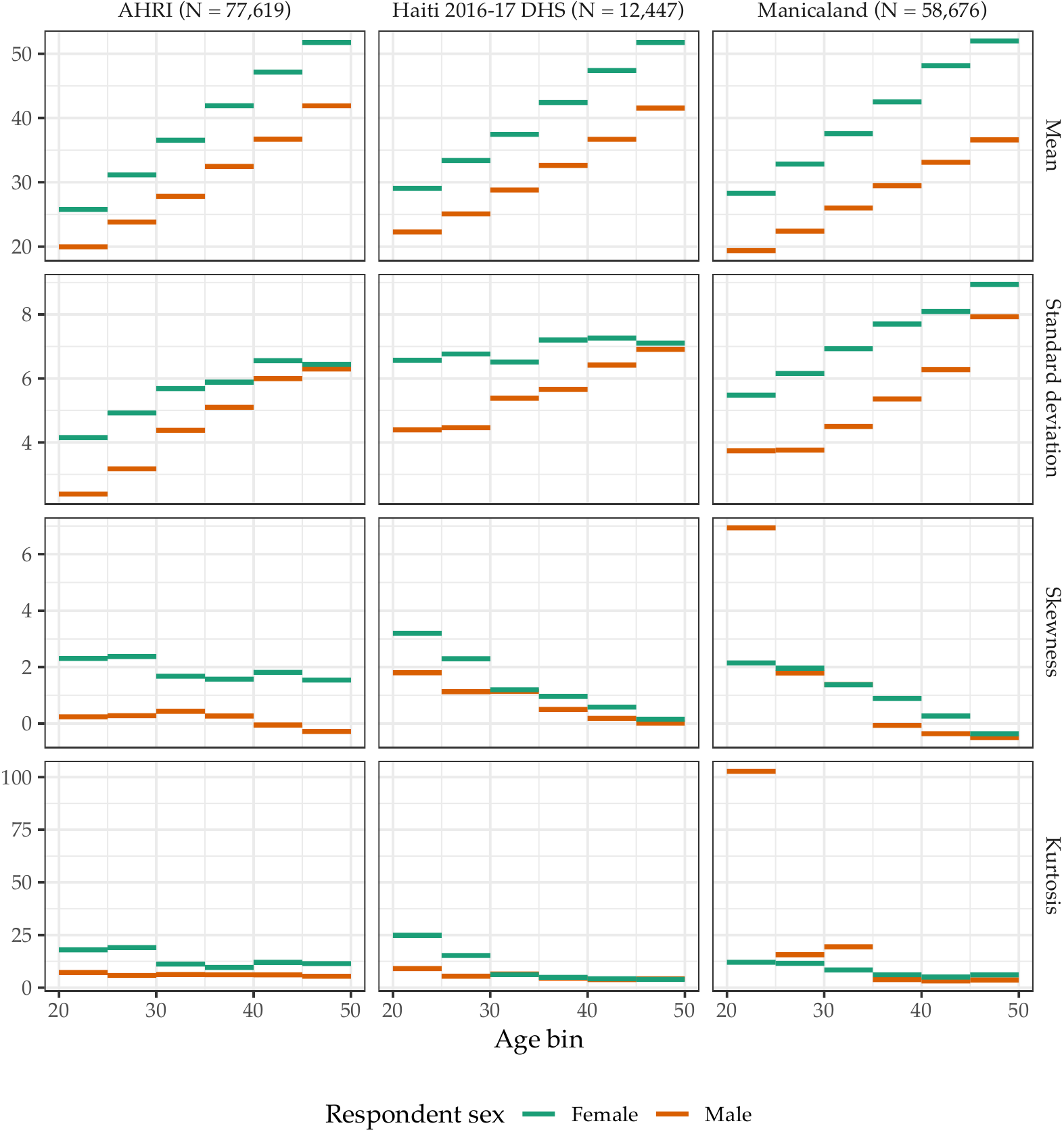} \caption{Observed means, variances, skewnesses, and kurtoses of partner age by five-year age bin and sex in all three data sets}\label{fig:meanPlot}
\end{figure}

The AHRI data included 77,619 partnerships, Manicaland had 58,676, and
the Haiti DHS had 12,447. As an illustrative example of the distribution
of partner ages, Figure \ref{fig:studyHist} presents histograms of
reported partner ages among women aged 35-39 for each of our three data
sets. Figure \ref{fig:meanPlot} shows the sex- and age bin-specific
empirical moments for the three data sets. Mean partner age increased
with respondent age consistently for both sexes across all three
data sets: among women, mean partner age increased by 26.0, 22.7, and
23.7 years in the AHRI data, Haiti DHS data, and Manicaland data,
respectively, between age bins 20-24 and 45-49. However, higher order
moments were less consistent: the standard deviation of women's
partners' ages changed by 2.3, 0.5, and 3.5 years in the AHRI data,
Haiti DHS data, and Manicaland data, respectively.

Within each data set, there is systematic variation across sex. For
example, the standard deviation of partner ages in the Haiti DHS
increased by 2.5 years among men and only by 0.5 years among women.
These summary statistics illustrate the heterogeneity of partner age
distributions across age and sex.

\hypertarget{probability-distribution-comparison-1}{%
\subsection{Probability distribution
comparison}\label{probability-distribution-comparison-1}}

\begin{figure}
\includegraphics[width=1\linewidth]{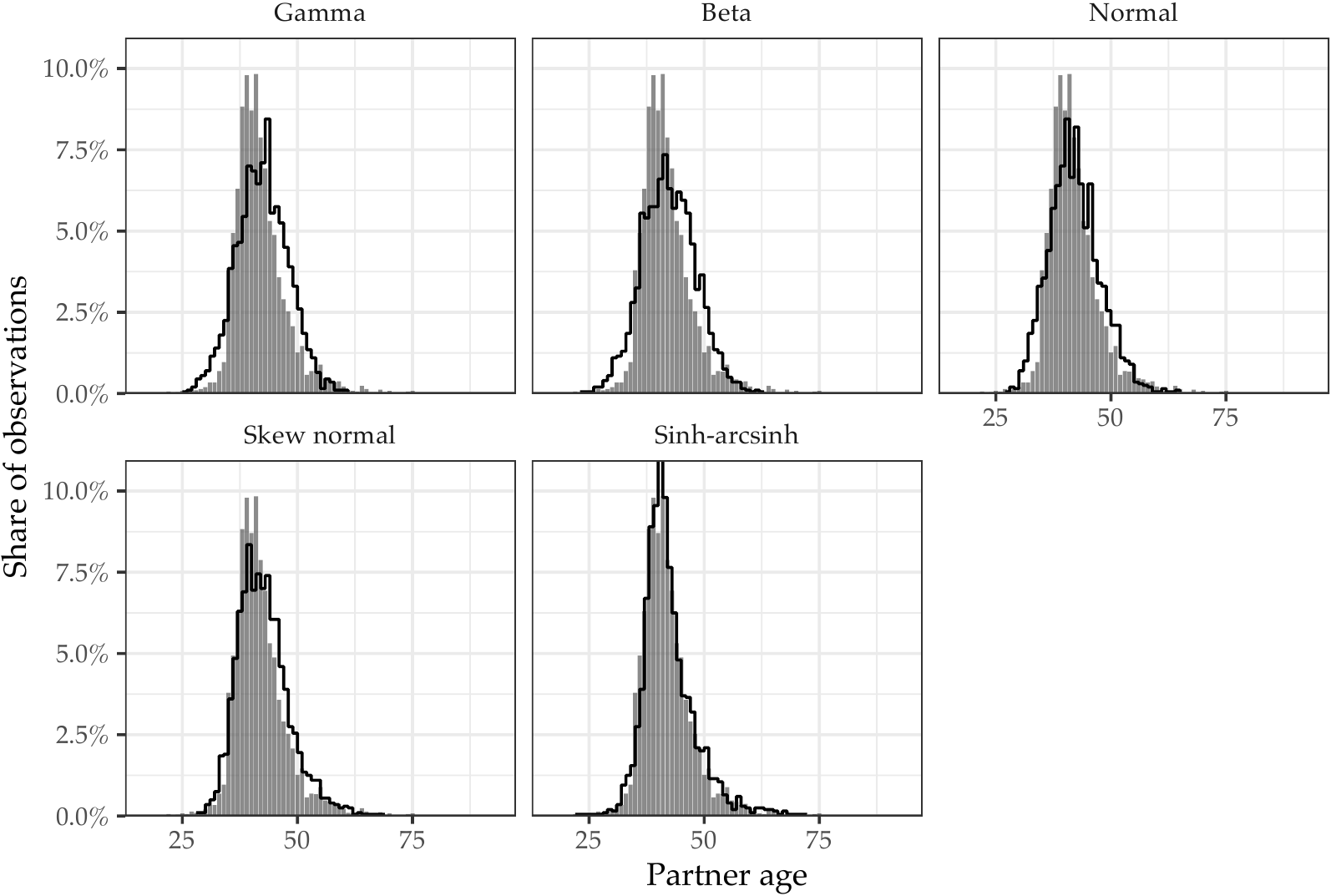} \caption{Observed partner age distributions (grey bars) and posterior predictive partner age  distributions (lines) for each probability distribution among women aged 35-39 in the AHRI data set. Posterior predictive distributions come from fitting each age bin/sex combination independently.}\label{fig:ahriWomen}
\end{figure}

To identify the probability distribution that most accurately described
the variation in sexual partner age distributions, we first determined the
dependent variable with the highest ELPD for each distribution-dependent
variable combination. Figure \ref{fig:ahriWomen} illustrates each
probability distribution's best fit to AHRI data among women aged 35-39
with each of the best distribution-specific dependent variables. Results
for all 36 data subsets and the 12 deheaped subsets are in Appendix
section ``Full results.''

The best dependent variable varied across data subset and probability
distribution. Table \ref{tab:depVarBestShare} provides the share of data
sets for which each dependent variable has the highest ELPD given each
distribution. The log-ratio dependent variable was best in 50.0\% of
subsets with a normal distribution, but it was best in only 27.8\% of
subsets with a skew normal distribution. The dependent variable that was
best in a plurality of subsets in each probability distribution
(i.e.~the variable with the highest percentage in each column of Table
\ref{tab:depVarBestShare}) used a \(\log\) link function. We restricted
all remaining comparisons to each distribution-subset combination's best
dependent variable.

\begin{table}

\caption{\label{tab:depVarBestShare}Share of subsets in which each dependent variable yields the highest ELPD given each probability distribution (excluding deheaped AHRI data).}
\centering
\begin{tabular}[t]{rrrr}
\toprule
Variable & Normal & Skew normal & Sinh-arcsinh\\
\midrule
Age difference & 22.2\% & 25.0\% & 16.7\%\\
Linear age & 8.3\% & 5.6\% & 16.7\%\\
Log-age & 19.4\% & 41.7\% & 30.6\%\\
Log-ratio & 50.0\% & 27.8\% & 36.1\%\\
\bottomrule
\end{tabular}
\end{table}

The sinh-arcsinh distribution had the highest ELPD in 35 of 36 data
subsets (98\%). In 29 of the 35 (83\%) cases in which the sinh-arcsinh
provided the highest ELPD, the absolute value of the ratio of the
difference between the two best ELPDs and the estimated standard error
of the difference was greater than 2, indicating that the sinh-arcsinh
distribution was significantly better than the alternatives in the
majority of cases. In one case, men aged 20-24 in the Haiti DHS, the
skew normal distribution resulted in a slightly higher ELPD than the
sinh-arcsinh distribution, but the standard error of the difference was
greater than the difference. These results were not affected by
deheaping the data (Appendix section ``Age heaping'').

To summarise each distribution's performance, we calculated the average
ELPD and QQ RMSE across the three data sets (Table
\ref{tab:likelihoodResultTab}). The sinh-arcsinh distribution had the
highest average ELPD and lowest average QQ RMSE in all three data sets.
The sinh-arcsinh distribution was, on average, able to predict the
empirical quantiles of each data set within half a year of accuracy
(0.36, 0.37, and 0.44 years for the AHRI, Haiti DHS, and Manicaland
data, respectively). Overlaid QQ plots and non-aggregated tables of ELPD differences and QQ RMSEs are presented in the ``Full results'' section of the Appendix.

\begin{table}

\caption{\label{tab:likelihoodResultTab}Model comparison metrics averaged across all data subsets for all three data sets. Higher ELPD values indicate better fit. Lower QQ RMSE values indicate more accurate prediction of empirical quantiles. Bolded rows are best across all three data sets.}
\centering
\begin{tabular}[t]{llll}
\toprule
Distribution & AHRI & Haiti 2016-17 DHS & Manicaland\\
\midrule
\addlinespace[0.3em]
\multicolumn{4}{l}{\textbf{ELPD}}\\
\hspace{1em}Gamma & \multicolumn{1}{r}{-14847.2} & \multicolumn{1}{r}{-2917.9} & \multicolumn{1}{r}{-13152.8}\\
\hspace{1em}Beta & \multicolumn{1}{r}{-14748.0} & \multicolumn{1}{r}{-2896.5} & \multicolumn{1}{r}{-13003.5}\\
\hspace{1em}Normal & \multicolumn{1}{r}{-14593.7} & \multicolumn{1}{r}{-2868.4} & \multicolumn{1}{r}{-12856.8}\\
\hspace{1em}Skew normal & \multicolumn{1}{r}{-14505.1} & \multicolumn{1}{r}{-2854.0} & \multicolumn{1}{r}{-12778.5}\\
\hspace{1em}Sinh-arcsinh & \multicolumn{1}{r}{\textbf{-14312.5}} & \multicolumn{1}{r}{\textbf{-2839.5}} & \multicolumn{1}{r}{\textbf{-12625.8}}\\
\addlinespace[0.3em]
\multicolumn{4}{l}{\textbf{QQ RMSE}}\\
\hspace{1em}Gamma & \multicolumn{1}{r}{0.83} & \multicolumn{1}{r}{0.82} & \multicolumn{1}{r}{0.95}\\
\hspace{1em}Beta & \multicolumn{1}{r}{0.99} & \multicolumn{1}{r}{0.82} & \multicolumn{1}{r}{1.11}\\
\hspace{1em}Normal & \multicolumn{1}{r}{0.82} & \multicolumn{1}{r}{0.68} & \multicolumn{1}{r}{0.97}\\
\hspace{1em}Skew normal & \multicolumn{1}{r}{0.77} & \multicolumn{1}{r}{0.65} & \multicolumn{1}{r}{0.85}\\
\hspace{1em}Sinh-arcsinh & \multicolumn{1}{r}{\textbf{0.36}} & \multicolumn{1}{r}{\textbf{0.37}} & \multicolumn{1}{r}{\textbf{0.44}}\\
\bottomrule
\end{tabular}
\end{table}

\hypertarget{distributional-regression-evaluation-1}{%
\subsection{Distributional regression
evaluation}\label{distributional-regression-evaluation-1}}

We fit all five distributional regression specifications to all three of
our data sets with sinh-arcsinh distributions and log-ratio dependent
variables and compared the ELPDs and QQ RMSEs as before (provided in
Table \ref{tab:allDistELPD}). Across all three data sets, the most
complex distributional model (Distributional 4) had the highest ELPD and
lowest QQ RMSE. When fit to the AHRI and Manicaland data sets (but not
for the Haiti DHS), the most complex distributional model was a least
two standard errors better than the next best model. Notably, the
largest ELPD improvements came from moving from conventional regression
(Conventional) to the simplest distributional model (improvements of 1646.0 units,
361.0 units, and 2181.2 units in the AHRI, Haiti DHS, and Manicaland
data, respectively). Full tables are available in the ``Full results''
section of the Appendix.

\begin{table}

\caption{\label{tab:allDistELPD}ELPD and QQ RMSE values for all five distributional regression models fit to each data set. The models increase in complexity from Conventional Regression to Distributional Model 4. Bolded ELPD values are more than two standard errors higher than the next best value in the column. Bolded QQ RMSE values are lowest in their column.}
\centering
\begin{tabular}[t]{llll}
\toprule
Model & AHRI & Haiti 2016-17 DHS & Manicaland\\
\midrule
\addlinespace[0.3em]
\multicolumn{4}{l}{\textbf{ELPD}}\\
\hspace{1em}Conventional & \multicolumn{1}{r}{52689.2} & \multicolumn{1}{r}{4777.8} & \multicolumn{1}{r}{21011.3}\\
\hspace{1em}Distributional 1 & \multicolumn{1}{r}{54335.2} & \multicolumn{1}{r}{5140.8} & \multicolumn{1}{r}{23192.5}\\
\hspace{1em}Distributional 2 & \multicolumn{1}{r}{54794.8} & \multicolumn{1}{r}{5138.7} & \multicolumn{1}{r}{23472.1}\\
\hspace{1em}Distributional 3 & \multicolumn{1}{r}{55534.2} & \multicolumn{1}{r}{5196.7} & \multicolumn{1}{r}{24313.7}\\
\hspace{1em}Distributional 4 & \multicolumn{1}{r}{\textbf{55841.9}} & \multicolumn{1}{r}{5207.6} & \multicolumn{1}{r}{\textbf{24516.1}}\\
\addlinespace[0.3em]
\multicolumn{4}{l}{\textbf{QQ RMSE}}\\
\hspace{1em}Conventional & \multicolumn{1}{r}{1.30} & \multicolumn{1}{r}{1.33} & \multicolumn{1}{r}{2.05}\\
\hspace{1em}Distributional 1 & \multicolumn{1}{r}{1.15} & \multicolumn{1}{r}{0.98} & \multicolumn{1}{r}{1.89}\\
\hspace{1em}Distributional 2 & \multicolumn{1}{r}{1.21} & \multicolumn{1}{r}{0.99} & \multicolumn{1}{r}{1.80}\\
\hspace{1em}Distributional 3 & \multicolumn{1}{r}{0.93} & \multicolumn{1}{r}{0.91} & \multicolumn{1}{r}{1.34}\\
\hspace{1em}Distributional 4 & \multicolumn{1}{r}{\textbf{0.66}} & \multicolumn{1}{r}{\textbf{0.84}} & \multicolumn{1}{r}{\textbf{1.04}}\\
\bottomrule
\end{tabular}
\end{table}

\begin{figure}
\includegraphics[width=1\linewidth]{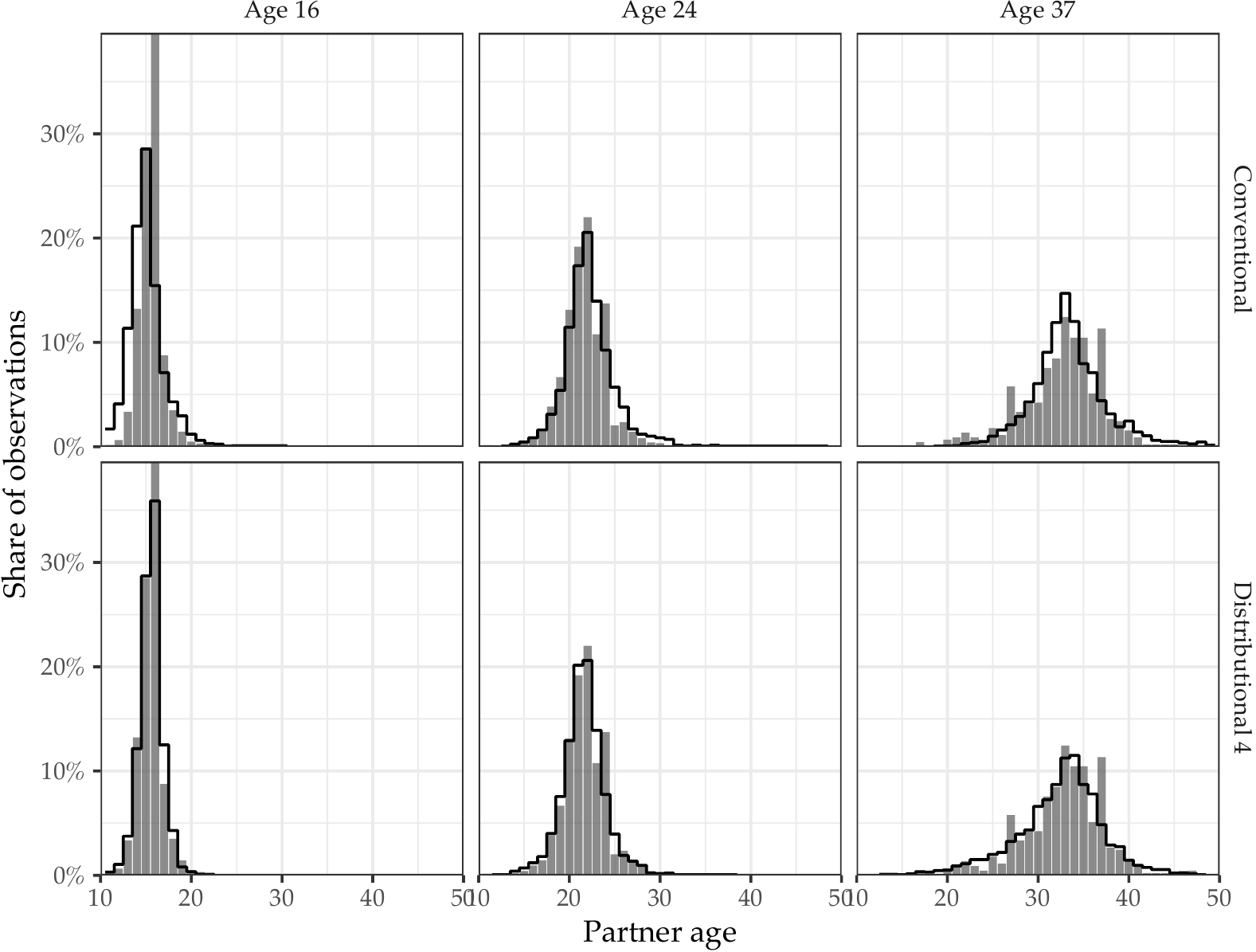} \caption{Observed partner age distributions (grey bars) and posterior predictive partner age  distributions (lines) for conventional regression and the most complex distributional model among men aged 16, 24, and 37 years in the AHRI data set. Posterior predictive distributions come from regression models fit to the entire AHRI data set.}\label{fig:AHRIpost}
\end{figure}

Figure \ref{fig:AHRIpost} shows the posterior predictive distributions
from the conventional regression model and the most complex
distributional model among men aged 16 years, 24 years, and 37 years in
the AHRI data to illustrate the effect of distributional regression. Not
only does the distributional model capture the high peak in the youngest
age more accurately, but it also allows the variance of the
distributions to change appropriately (beyond the change that naturally
results from the log link function).

\begin{figure}
\includegraphics[width=1\linewidth]{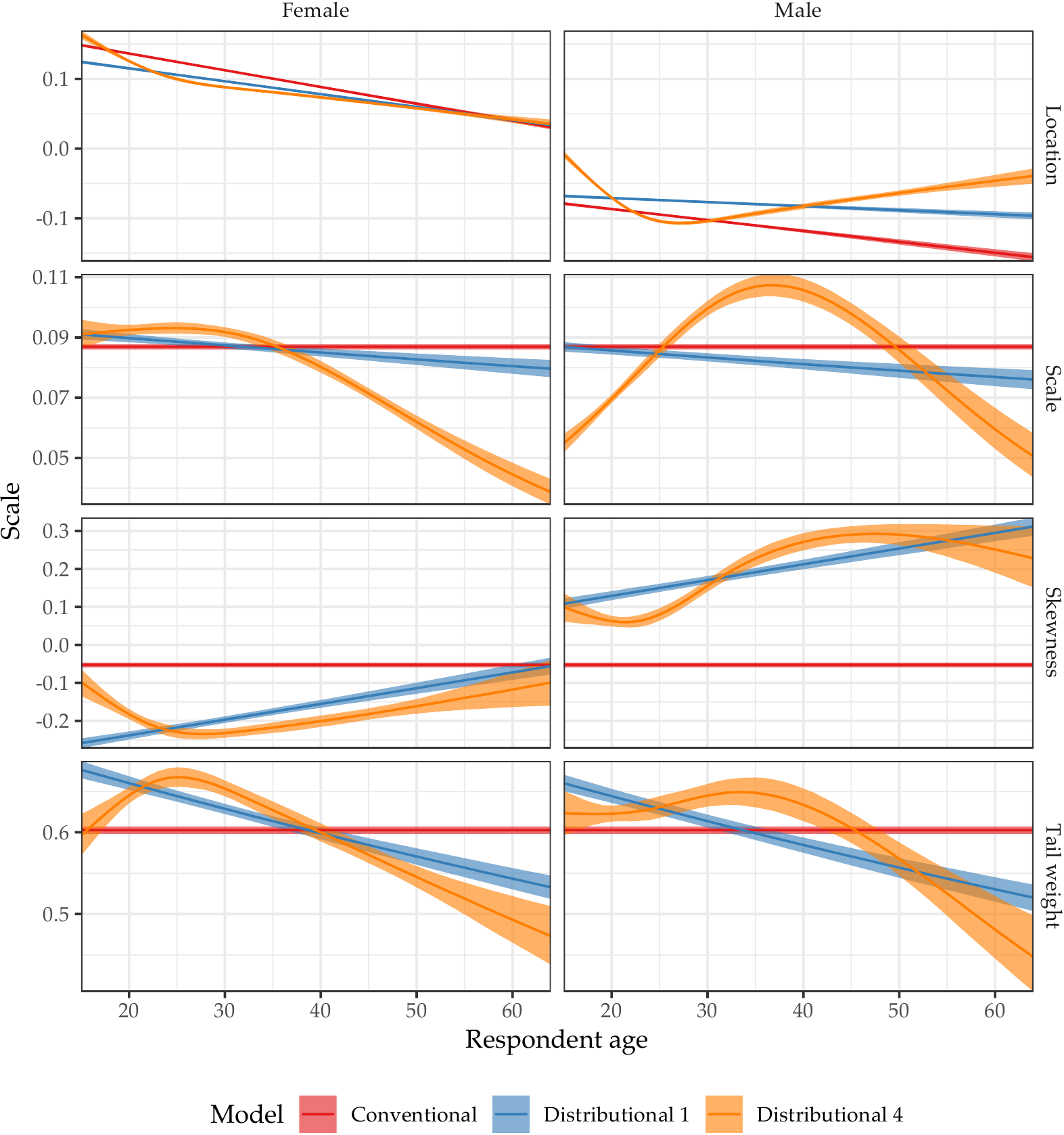} \caption{Estimated sinh-arcsinh distributional parameters from the conventional regression model, and distributional models 1 and 4 fit to the AHRI data. “Conventional" assumes no variation across age and sex, “Distributional 1" allows for independent age and sex effects, and “Distributional 4" includes sex-specific splines with respect to age.}\label{fig:summaryComparison}
\end{figure}

Figure \ref{fig:summaryComparison} illustrates posterior summaries among
men and women in the AHRI data for all four distributional parameters
for the conventional regression model, the simplest distributional
model, and the most complex distributional model. The red estimates
(Conventional Regression) of the three higher order parameters were
constant across age and sex, whereas the blue estimates (Distributional
Model 1) included independent, linear age and sex effects. The orange
estimates (Distributional Model 4) were generated sex-specific splines
with respect to age, allowing for flexible variation across age and sex.

The third row of plots in Figure \ref{fig:summaryComparison}, which
corresponds to the skewness parameter, illustrates the impact of
incorporating sex and age effects into the model. The conventional
regression model estimated that neither the distrbution for men nor
women exhibited much skewness; the estimated parameter value was -0.05
(95\% UI: -0.06 to -0.05) regardless of age, with \(0.0\) corresponding
to perfect symmetry. However, when we allowed independent age and sex
effects in Distributional Model 1, we estimated that at age 15, women's
skewness was -0.26 (95\% UI: -0.27 to -0.25) and men's was 0.11 (95\%
UI: 0.10 to 0.12).

The most complex model (Distributional Model 4) inferred sex-specific,
non-linear variation with respect to age in all four distributional
parameters. The non-linearity was particularly dramatic in the scale
parameter among men. The scale value began at 0.05 (95\% UI: 0.05 to
0.06) among 15-year-olds, peaked among 37-year-olds at 0.11 (95\% UI:
0.10 to 0.11), and decreased back down to 0.05 (95\% UI: 0.04 to 0.06)
at age 64.

\begin{figure}
\includegraphics[width=1\linewidth]{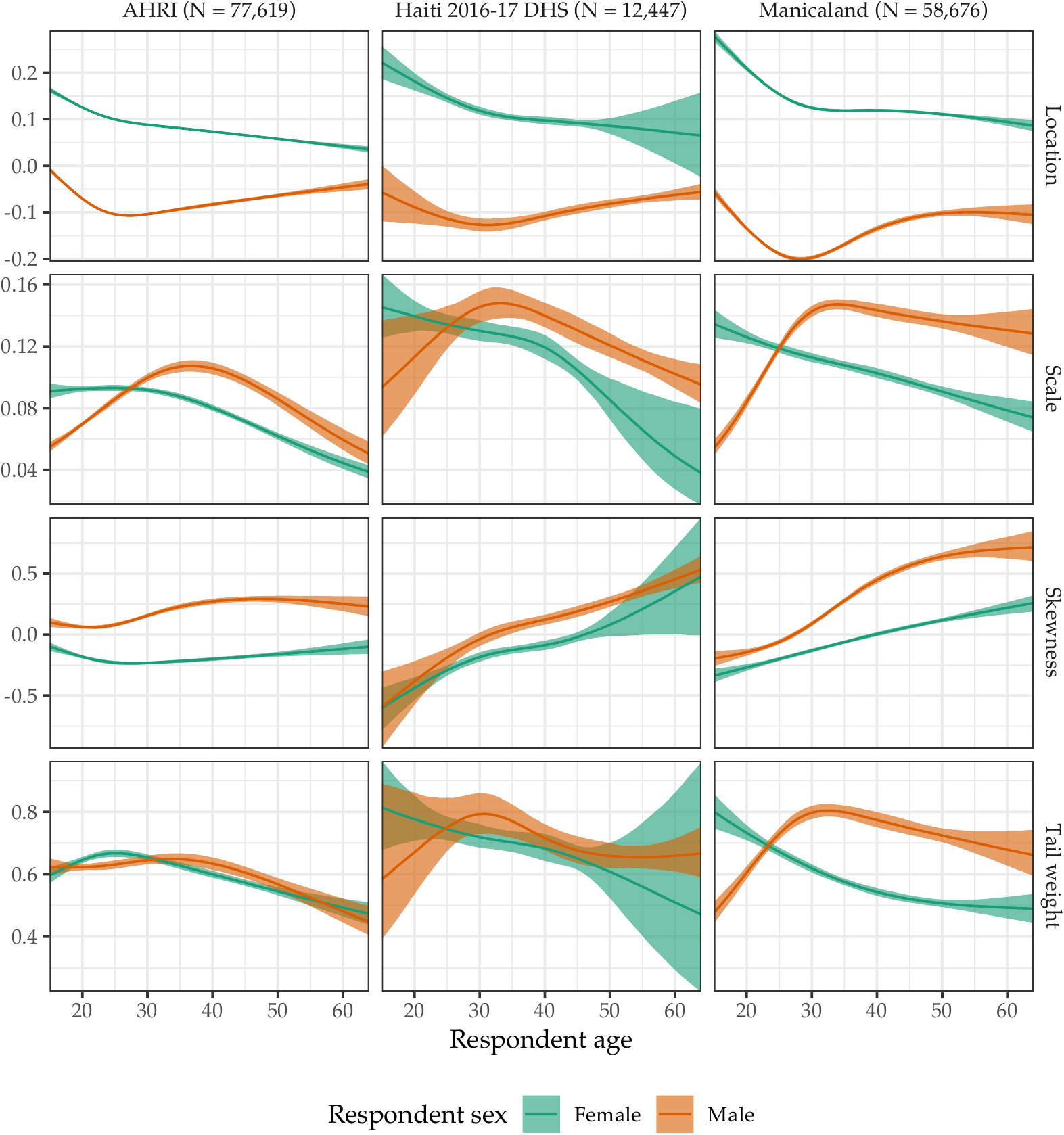} \caption{Estimated sinh-arcsinh distributional parameters for Distributional Model 4 fit to the three main data sets.}\label{fig:modelAllMoments}
\end{figure}

Finally, Figure \ref{fig:modelAllMoments} presents inferred
distributional parameters from Distributional Model 4 for both men and
women for all three data sets. Based on those plots, the flexible model
was justified for most distributional parameters in all three data sets.
Were we to continue developing these models, this plot suggests that
skewness might only need linear, sex-specific effects with respect to
age. Interestingly, the 2016-2017 Haiti DHS and Manicaland estimates
exhibit similar patterns across all four parameters, despite the
different socio-cultural contexts surrounding partnerships in the two
populations. We also note that the DHS does not collect data on adults
aged 50 years and older, so our estimates in Haiti from age 50 to age 64
are purely extrapolated.

\hypertarget{discussion}{%
\section{Discussion}\label{discussion}}

We found that the sinh-arcsinh distribution reproduced observed sexual
partner age distributions better than a number of other possible
distributional assumptions across age and sex in three distinct data
sets. We integrated this finding into a distributional regression
framework using existing statistical modelling software. Even the
simplest distributional regression in our set of candidate models far
outperformed conventional regression, in which all moments except the
first are estimated as constants. Our most complex distributional model
fit better than all other models in all three data sets, suggesting that
modelling these data benefits from the additional complexity.

These results indicate that distributional regression models with
sinh-arcsinh distributions can accurately replicate age-/sex-specific
sexual partner age distributions. This approach presents a number of
advantages over previous methods. First, like Smid et al., it allows a
unique distribution for every age-sex combination. As Figure
\ref{fig:meanPlot} illustrates, partner age distributions can exhibit
substantial, systematic variation across age and sex in any of the first
four moments, so we must consider modelling strategies that allow for
such variation. Second, distributional regression offers a principled
method to propagate uncertainty through this estimation process.

Finally, distributional regression implemented through \texttt{brms}
provides access to a deep set of hierarchical modelling tools that could
enable estimation in a variety of low-data settings. We evaluated a
small set of relatively simple distributional models in this work, but,
theoretically, each distributional parameter could have its own,
arbitrarily complex hierarchical regression model. Using these tools,
one could estimate unique partner age distributions across levels of
stratification that are substantively interesting but do not provide
sufficient sample size for independent estimation (e.g.~study sites or
geographic areas).

We have identified several limitations in this approach. First, the
amount of data required to produce usefully precise estimates is not
tested. Each additional distributional parameter introduces model
parameters, so this method is more complex than conventional regression.
The sinh-arcsinh distribution did fail to produce the highest ELPD in
our smallest data subset (N = 170), but it was not significantly worse
than the best distribution. More importantly, by integrating these data
into a distributional modelling framework, we gain the ability to impose
structure on these parameters, which could easily offset the cost of any
additional model parameters.

Interpreting the inferred model parameters in sinh-arcsinh regression
can also be difficult. Whereas conventional regression estimates the
effects of covariates on expected values, the sinh-arcsinh distribution
is parametrised in terms of a location parameter. This parameter
correlates closely with the central tendency of the distribution, but it
is not strictly equal to the mean. We can reparametrise the distribution so that
we estimate a mean (and therefore effects of covariates on the expected
value), but it is not currently possible in the probabilistic
programming software that underlies \texttt{brms}.

Third, our analysis assumed that we were operating at a level of
stratification at which partnerships are basically comparable, but any
number of factors could lead to fundamentally different partner age
distributions. For example, we did not control for whether the
partnership was same-sex or the type of the partnership (married,
casual, etc.). That said, our distributional framework would allow us to
incorporate data on any of those factors directly into the model.

Despite these limitations, we believe that the strategy we present will
work well in future projects that require estimates of partner age
distributions. We plan to use these methods to produce age-mixing
matrices to inform epidemic models of HIV, but there are a number of
additional directions that could be explored. We are specifically interested in
leveraging the spatio-temporal structure of the survey data used here.
Hierarchical mapping exercises with household survey data are
increasingly common in epidemiology, but estimating spatially varying
partner age distributions would require an evaluation of how best to
model higher order moments over space. We would, for example, need to
consider how the variance of partner age distributions varies by
urbanicity.

Similarly, population-based studies typically collect far more detailed
information on partnerships than we took advantage of here. Relationship
type is a key confounder of the association between respondent age and
partner age (that we ignored for the purposes of our experiments). We
might expect the age distributions of casual partners to
vary substantially from those of long-term cohabiting partners. Because
we have built our model in a pre-existing regression framework,
incorporating new covariates into any of the distributional regression
specifications should be simple.

We believe that our framework offers a flexible, accurate, and robust
method for smoothing and interpolating sexual partner age distributions,
but these methods are not specific to partner age distributions. The
sinh-arcsinh distribution is relatively easy to implement without incurring
high computational cost, so it could be applied in many settings.
Even without the distributional regression framework we have used here,
allowing the third and fourth moments of the distribution to vary from
the ``default'' normal values could be valuable across a variety of
applications.

Distributional regression is also underutilised in social science
applications. We often work with large surveys that would comfortably
support models for higher order parameters. Data requirements will vary
by application and model, but, as we have shown here, even a simple
distributional model can improve fit and avoid biasing estimates.

\hypertarget{acknowledgements}{%
\section{Acknowledgements}\label{acknowledgements}}

JWE, SG, and KAR acknowledge funding support from the Bill \& Melinda Gates
Foundation. JWE and SG acknowledge funding support from the MRC Centre for Global Infectious Disease Analysis
(reference MR/R015600/1), jointly funded by the UK Medical Research
Council (MRC) and the UK Foreign, Commonwealth \& Development Office
(FCDO), under the MRC/FCDO Concordat agreement and is also part of the
EDCTP2 programme supported by the European Union. JWE also acknowledges
funding support from National Institute of Allergy and Infectious
Disease of the National Institutes of Health under award number
R01AI136664. The content is solely the responsibility of the authors and
does not necessarily represent the official views of the National
Institutes of Health. SRF acknowledges funding support from the EPSRC
(EP/V002910/1). TMW's work is funded by the Imperial College President's
PhD Scholarship. We thank all of the people who participated in the three studies that shared their data with us, as well as the survey and data management teams, without whom this work would not be possible.

\hypertarget{competing-interests}{%
\section{Competing interests}\label{competing-interests}}

None declared.

\hypertarget{references}{%
\section*{References}\label{references}}
\addcontentsline{toc}{section}{References}
\singlespacing
\hypertarget{refs}{}
\begin{CSLReferences}{1}{0}
\leavevmode\hypertarget{ref-anderson_age-dependent_1992}{}%
Anderson, R. M., May, R. M., Ng, T. W., \& Rowley, J. T. (1992).
Age-dependent choice of sexual partners and the transmission dynamics of
{HIV} in sub-saharan africa. \emph{Philosophical Transactions of the
Royal Society of London. Series B: Biological Sciences},
\emph{336}(1277), 135--155. \url{https://doi.org/10.1098/rstb.1992.0052}

\leavevmode\hypertarget{ref-beauclair_role_2018}{}%
Beauclair, R., Hens, N., \& Delva, W. (2018). The role of age-mixing
patterns in {HIV} transmission dynamics: Novel hypotheses from a field
study in cape town, south africa. \emph{Epidemics}, \emph{25}, 61--71.
\url{https://doi.org/10.1016/j.epidem.2018.05.006}

\leavevmode\hypertarget{ref-burkner_advanced_2018}{}%
Bürkner, P.-C. (2018). Advanced bayesian multilevel modeling with the r
package brms. \emph{The R Journal}, \emph{10}(1), 395--411.
\url{https://doi.org/10.32614/RJ-2018-017}

\leavevmode\hypertarget{ref-gareta_cohort_2021}{}%
Gareta, D., Baisley, K., Mngomezulu, T., Smit, T., Khoza, T., Nxumalo,
S., Dreyer, J., Dube, S., Majozi, N., Ording-Jesperson, G., Ehlers, E.,
Harling, G., Shahmanesh, M., Siedner, M., Hanekom, W., \& Herbst, K.
(2021). Cohort profile update: Africa centre demographic information
system ({ACDIS}) and population-based {HIV} survey. \emph{International
Journal of Epidemiology}, \emph{50}(1), 33--34.
\url{https://doi.org/10.1093/ije/dyaa264}

\leavevmode\hypertarget{ref-gareta_ahripipmens_2020}{}%
Gareta, D., Dube, S., \& Herbst, K. (2020a). \emph{{AHRI}.{PIP}.men's
general health.all.release 2020-07}. Africa Health Research Institute
({AHRI}). \url{https://doi.org/10.23664/AHRI.PIP.RD04-99.MGH.ALL.202007}

\leavevmode\hypertarget{ref-gareta_ahripipwomens_2020}{}%
Gareta, D., Dube, S., \& Herbst, K. (2020b). \emph{{AHRI}.{PIP}.women's
general health.all.release 2020-07}. Africa Health Research Institute
({AHRI}). \url{https://doi.org/10.23664/AHRI.PIP.RD03-99.WGH.ALL.202007}

\leavevmode\hypertarget{ref-garnett_balancing_1994}{}%
Garnett, G. P., \& Anderson, R. M. (1994). Balancing sexual partnerships
in an age and activity stratified model of {HIV} transmission in
heterosexual populations. \emph{{IMA} Journal of Mathematics Applied in
Medicine and Biology}, \emph{11}(3), 161--192.
\url{https://doi.org/10.1093/imammb/11.3.161}

\leavevmode\hypertarget{ref-gregson_documenting_2017}{}%
Gregson, S., Mugurungi, O., Eaton, J., Takaruza, A., Rhead, R., Maswera,
R., Mutsvangwa, J., Mayini, J., Skovdal, M., Schaefer, R., Hallett, T.,
Sherr, L., Munyati, S., Mason, P., Campbell, C., Garnett, G. P., \&
Nyamukapa, C. A. (2017). Documenting and explaining the {HIV} decline in
east zimbabwe: The manicaland general population cohort. \emph{{BMJ}
Open}, \emph{7}(10), e015898.
\url{https://doi.org/10.1136/bmjopen-2017-015898}

\leavevmode\hypertarget{ref-gregson_sexual_2002}{}%
Gregson, S., Nyamukapa, C. A., Garnett, G. P., Mason, P. R., Zhuwau, T.,
Caraël, M., Chandiwana, S. K., \& Anderson, R. M. (2002). Sexual mixing
patterns and sex-differentials in teenage exposure to {HIV} infection in
rural zimbabwe. \emph{Lancet (London, England)}, \emph{359}(9321),
1896--1903. \url{https://doi.org/10.1016/S0140-6736(02)08780-9}

\leavevmode\hypertarget{ref-hallett_behaviour_2007}{}%
Hallett, T. B., Gregson, S., Lewis, J. J. C., Lopman, B. A., \& Garnett,
G. P. (2007). Behaviour change in generalised {HIV} epidemics: Impact of
reducing cross-generational sex and delaying age at sexual debut.
\emph{Sexually Transmitted Infections}, \emph{83 Suppl 1}, i50--54.
\url{https://doi.org/10.1136/sti.2006.023606}

\leavevmode\hypertarget{ref-harling_age-disparate_2014}{}%
Harling, G., Newell, M.-L., Tanser, F., Kawachi, I., Subramanian, S., \&
Bärnighausen, T. (2014). Do age-disparate relationships drive {HIV}
incidence in young women? Evidence from a population cohort in rural
{KwaZulu}-natal, south africa. \emph{Journal of Acquired Immune
Deficiency Syndromes (1999)}, \emph{66}(4), 443--451.
\url{https://doi.org/10.1097/QAI.0000000000000198}

\leavevmode\hypertarget{ref-institut_haitien_de_lenfance_-_ihehaiti_haiti_2018}{}%
Institut Haïtien de l'Enfance - IHE/Haiti, \& ICF. (2018). \emph{Haiti
enquête mortalité, morbidité et utilisation des services 2016-2017 -
{EMMUS}-{VI} {[}dataset{]}}. {IHE}/Haiti, {ICF} {[}Producers{]}.
\url{http://dhsprogram.com/pubs/pdf/FR326/FR326.pdf}

\leavevmode\hypertarget{ref-johnson_systems_1949}{}%
Johnson, N. L. (1949). Systems of frequency curves generated by methods
of translation. \emph{Biometrika}, \emph{36}(1), 149.
\url{https://doi.org/10.2307/2332539}

\leavevmode\hypertarget{ref-jones_sinh-arcsinh_2009}{}%
Jones, M. C., \& Pewsey, A. (2009). Sinh-arcsinh distributions.
\emph{Biometrika}, \emph{96}(4), 761--780.
\url{https://doi.org/10.1093/biomet/asp053}

\leavevmode\hypertarget{ref-kneib_primer_2017}{}%
Kneib, T., \& Umlauf, N. (2017). \emph{A primer on bayesian
distributional regression} (Working Paper No. 2017-13). Working Papers
in Economics; Statistics.
\url{https://www.econstor.eu/handle/10419/180164}

\leavevmode\hypertarget{ref-maughan-brown_sexual_2016}{}%
Maughan-Brown, B., Evans, M., \& George, G. (2016). Sexual behaviour of
men and women within age-disparate partnerships in south africa:
Implications for young women's {HIV} risk. \emph{{PLOS} {ONE}},
\emph{11}(8), e0159162.
\url{https://doi.org/10.1371/journal.pone.0159162}

\leavevmode\hypertarget{ref-r_core_team_r_2020}{}%
R Core Team. (2020). \emph{R: A language and environment for statistical
computing}. R Foundation for Statistical Computing.
\url{https://www.R-project.org}

\leavevmode\hypertarget{ref-reniers_data_2016}{}%
Reniers, G., Wamukoya, M., Urassa, M., Nyaguara, A., Nakiyingi-Miiro,
J., Lutalo, T., Hosegood, V., Gregson, S., Gómez-Olivé, X., Geubbels,
E., Crampin, A. C., Wringe, A., Waswa, L., Tollman, S., Todd, J.,
Slaymaker, E., Serwadda, D., Price, A., Oti, S., \ldots{} Zaba, B.
(2016). Data resource profile: Network for analysing longitudinal
population-based {HIV}/{AIDS} data on africa ({ALPHA} network).
\emph{International Journal of Epidemiology}, \emph{45}(1), 83--93.
\url{https://doi.org/10.1093/ije/dyv343}

\leavevmode\hypertarget{ref-ritchwood_characteristics_2016}{}%
Ritchwood, T. D., Hughes, J. P., Jennings, L., MacPhail, C., Williamson,
B., Selin, A., Kathleen, K., Gómez-Olivé, F. X., \& Pettifor, A. (2016).
Characteristics of age-discordant partnerships associated with {HIV}
risk among young south african women ({HPTN} 068). \emph{Journal of
Acquired Immune Deficiency Syndromes (1999)}, \emph{72}(4), 423--429.
\url{https://doi.org/10.1097/QAI.0000000000000988}

\leavevmode\hypertarget{ref-smid_age_2018}{}%
Smid, J. H., Garcia, V., Low, N., Mercer, C. H., \& Althaus, C. L.
(2018). Age difference between heterosexual partners in britain:
Implications for the spread of chlamydia trachomatis. \emph{Epidemics},
\emph{24}, 60--66. \url{https://doi.org/10.1016/j.epidem.2018.03.004}

\leavevmode\hypertarget{ref-noauthor_dhs_2021}{}%
\emph{The {DHS} program}. (2021). {USAID}.
\url{https://dhsprogram.com/data/}

\leavevmode\hypertarget{ref-vehtari_loo_2020}{}%
Vehtari, A., Gabry, J., Magnusson, M., Yao, Y., Bürkner, P.-C.,
Paananen, T., \& Gelman, A. (2020). \emph{Loo: Efficient leave-one-out
cross-validation and {WAIC} for bayesian models}.
\url{https://mc-stan.org/loo}

\leavevmode\hypertarget{ref-vehtari_practical_2017}{}%
Vehtari, A., Gelman, A., \& Gabry, J. (2017). Practical bayesian model
evaluation using leave-one-out cross-validation and {WAIC}.
\emph{Statistics and Computing}, \emph{27}(5), 1413--1432.
\url{https://doi.org/10.1007/s11222-016-9696-4}

\leavevmode\hypertarget{ref-wickham_ggplot2_2016}{}%
Wickham, H. (2016). \emph{ggplot2: Elegant graphics for data analysis}.
Springer-Verlag New York. \url{https://ggplot2.tidyverse.org}

\end{CSLReferences}

\newpage

\hypertarget{appendix}{%
\section{Appendix}\label{appendix}}
\onehalfspacing
\hypertarget{heaping}{%
\subsection{Age heaping}\label{heaping}}

\begin{figure}
\includegraphics[width=1\linewidth]{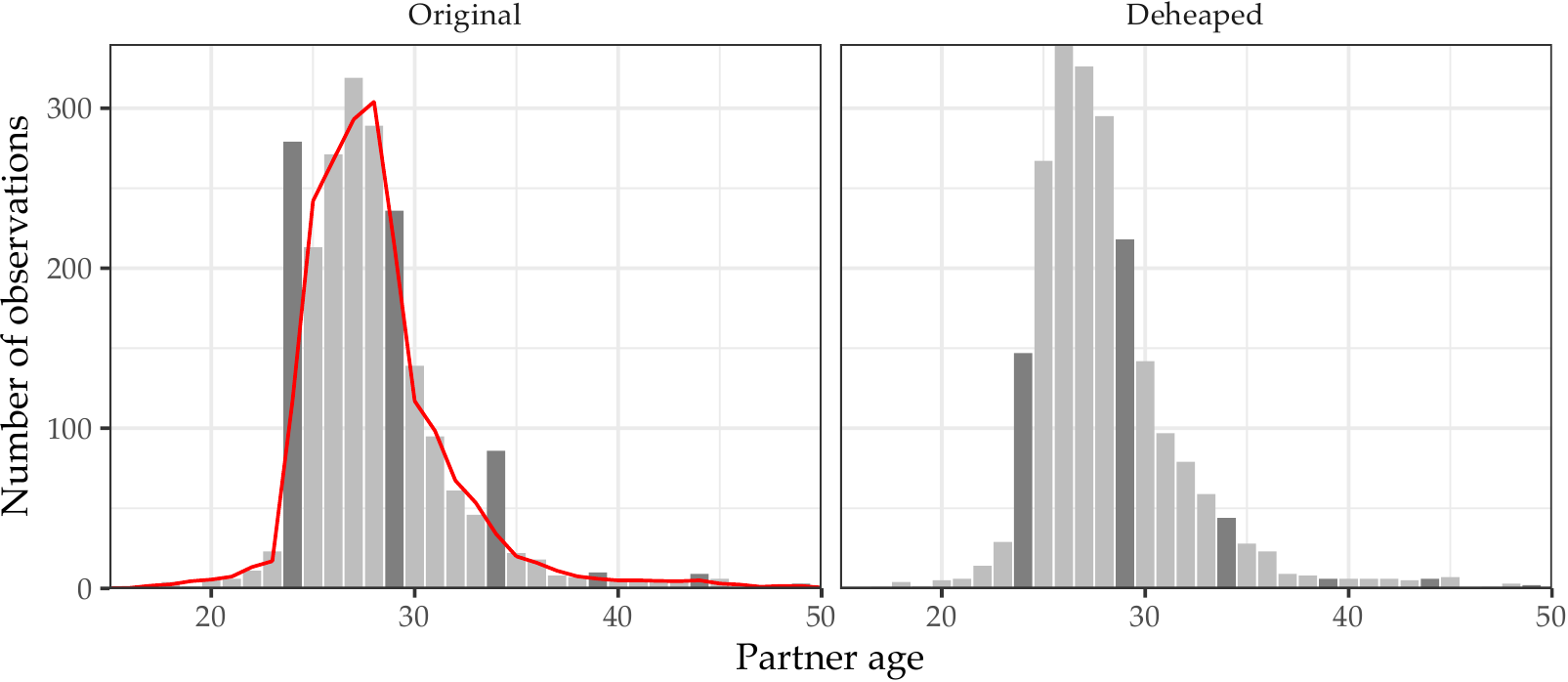} \caption{Illustration of the effect of the deheaping algorithm on women aged exactly 24 years in the AHRI data. Dark grey bars correspond to ages identified as potentially heaped (multiples of five away from 24). The red line is the expected count of observations estimated by excluding any potentially heaped ages.}\label{fig:heapPlot}
\end{figure}

Respondents in each of these data sets are disproportionately likely to
report that their partners' ages are multiples of five or multiples of
five away from their own age, leading to distinct ``spikes'' in the
empirical partner age (or age difference) distributions at multiples of
five. The left panel of Figure \ref{fig:heapPlot} illustrates this
phenomenon among women aged 24 years in the AHRI data. These spikes,
widely referred to as ``heaping,'' could bias our results towards
certain probability distributions, so we developed a simple deheaping
algorithm, applied it to the AHRI data.

To account for the possibility that heaping affected the results, we
developed a simple deheaping algorithm and treated the deheaped AHRI
data as a fourth data set. Due to the structure of the questionnaire
(``how many years older or younger is your partner than you?''), the AHRI
partner age data exhibit strong heaping on partner ages that are
multiples of five years from the respondent's age. For example, among
women aged 24 years, we observe far more partners aged exactly 29 years
than expected.

Let \(n_{s,a,p}\) be the number of observed partnerships with \(s_i=s\),
\(a_i=a\), and \(p_i=p\). Fixing age to be \(a\) and sex to be \(s\), we
can find the expected count at partner age \(p\), \(\hat n_{s,a,p}\) by
fitting a Nadaraya-Watson estimator to all ordered pairs
\((p, n_{s,a,p})\) such that \(p-a\) is \emph{not} a multiple of five.
We can then find the positive-valued excess counts at all \(p\) such
that \(p-a\) \emph{is} a multiple of five:

\[
e_{s,a,p} = \max(n_{s,a,p}-\hat n_{s,a,p}, 0).
\]

This quantity, \(e_{s,a,p}\), is what the Nadaraya-Watson estimator has
identified as number of heaped observations. Fixing \(p^\star\) to be a
partner age such that \((p^\star-a)\bmod5\equiv0\), we assume that all
of the excess mass at \(p^\star\) will be allocated to the four partner
ages on either side of \(p^\star\). We find the share of
\(e_{s,a,p^\star}\) to be allocated to each of
\((p^\star-2,.. .,p^\star+2)\), denoted \(b_{s,a, p}\), as

\[
b_{s,a, p} =\frac{n_{s,a,p}}{\sum_{i=-2}^2n_{s,a,p^\star+i}},
\]

substituting in \(\hat n_{s,a,p^\star}\) for \(n_{s,a,p}\) wherever
applicable. Finally, we find the number of individuals to be reassigned
from \(p ^\star\) to each \(p\) within two years of \(y^\star\) as
\(d_{s,a,p}=b_{s,a,p}\cdot e_{s,a,p^\star}\). Note that each partner age
can only ``receive'' partnerships from its nearest multiple of five and
that each multiple of five can only ``send'' partnerships to itself and
the four partner ages on either side of it. For each \(y\) within two
years of \(y^\star\), we randomly select \(\lfloor d_{s,a,p}\rceil\)
individuals to move from \(p^\star\) to \(p\). We apply this method for
both sexes and all respondent ages with at least two observations
separately.

Figure \ref{fig:heapPlot} illustrates the effect of this process on data
among women aged 24 in the AHRI data. Despite its simplicity, the
deheaping algorithm seems to produce distributions that should be
sufficiently plausible for the purposes of this experiment. If our
results differed substantially between the original and deheaped AHRI
data, we would need to consider the possibility that our results could
be an artefact of heaping.

This method is quite simple, but it seems to work reasonably well on the
AHRI data. Regardless, we do not need a perfect deheaping algorithm for
this application; we just need one that will give us a \emph{plausibly}
deheaped version of the AHRI data. If the results differ drastically
between the heaped and deheaped data sets (\emph{i.e.} if one probability
distribution works perfectly only on the deheaped data), then we will
know that our results are sensitive to irregularities in the data.

\hypertarget{results-1}{%
\subsubsection{Results}\label{results-1}}

\begin{figure}
\includegraphics[width=1\linewidth]{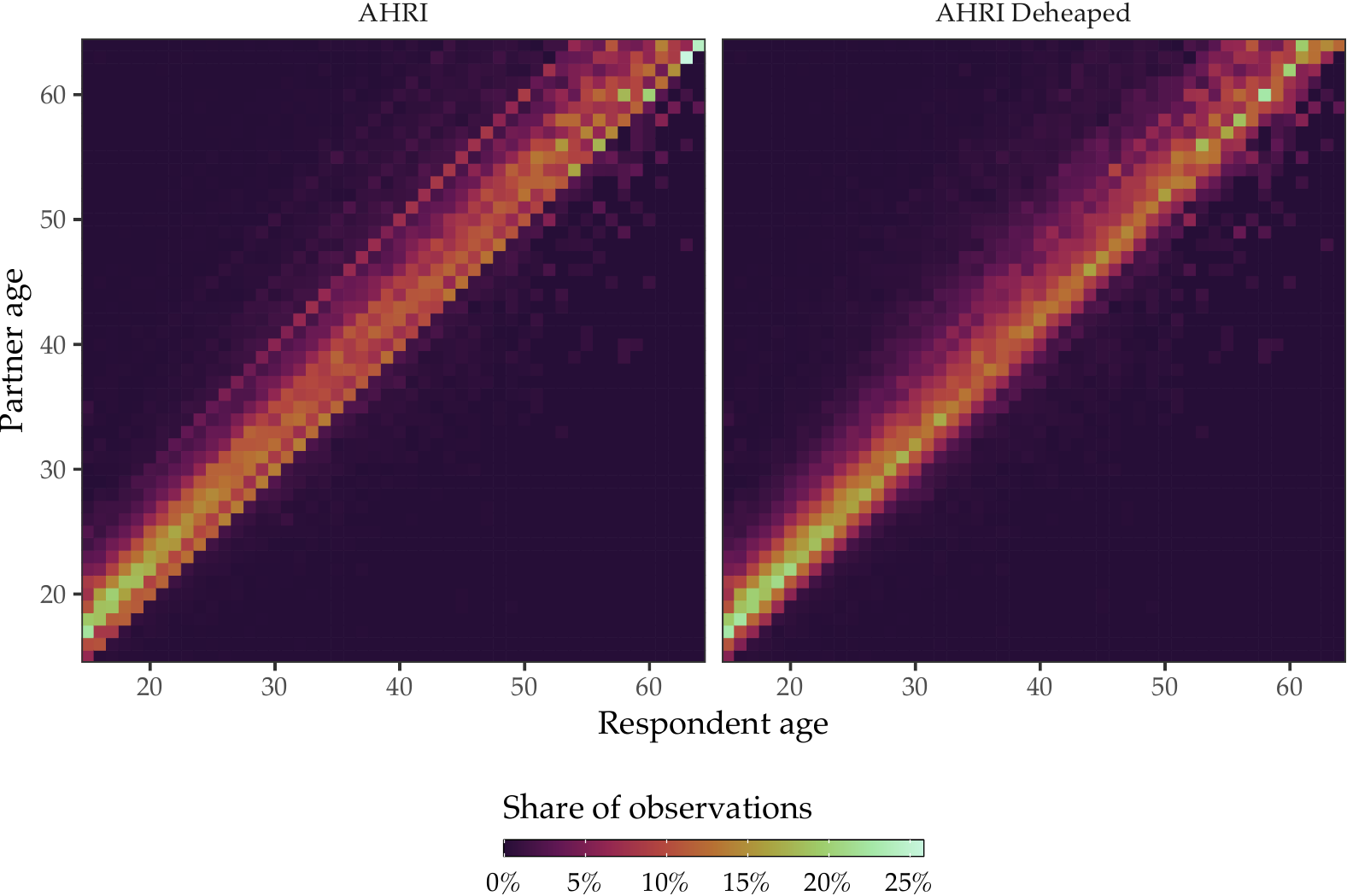} \caption{Observed sexual partner age distributions among women in the AHRI data. The left panel is original data, and the right panel is the same data set after deheaping age differences from multiples of five.}\label{fig:deheapPlot}
\end{figure}

Figure \ref{fig:deheapPlot} shows the presence of age heaping among
women in the AHRI data, as well as the effects of our deheaping
algorithm. Visible diagonal lines indicate that women were
disproportionately likely to report that the difference between their
partner's age and their own age was a multiple of five. Heaping to
partner ages (not partner age differences) would manifest as horizontal
lines. As we can see in the right panel, the deheaping procedure
resolves the majority of the heaping. We cannot validate the algorithm,
but for the purposes of this experiment, simply producing plausibly
deheaped age distributions should be sufficient.

Table \ref{tab:deheapTab} provides ELPD and QQ RMSE values for all five
regression models fit to the deheaped AHRI data. As with the heaped AHRI
data, the most complex distributional model had the highest ELPD
(58504.0). From these results, we conclude that the presence of heaping
in the three main data sets is unlikely to have substantially altered the
results of this analysis.

\begin{table}

\caption{\label{tab:deheapTab}ELPD and QQ RMSE values for all five models fit to deheaped AHRI data The models increase in complexity from Conventional Regression to Distributional Model 4. Bolded ELPD values are more than two standard errors higher than the next best value in the column. Bolded QQ RMSE values are lowest in their column.}
\centering
\begin{tabular}[t]{ll}
\toprule
Model & AHRI Deheaped\\
\midrule
\addlinespace[0.3em]
\multicolumn{2}{l}{\textbf{ELPD}}\\
\hspace{1em}Conventional & \multicolumn{1}{r}{55296.2}\\
\hspace{1em}Distributional 1 & \multicolumn{1}{r}{57097.4}\\
\hspace{1em}Distributional 2 & \multicolumn{1}{r}{57503.7}\\
\hspace{1em}Distributional 3 & \multicolumn{1}{r}{58219.2}\\
\hspace{1em}Distributional 4 & \multicolumn{1}{r}{\textbf{58504.0}}\\
\addlinespace[0.3em]
\multicolumn{2}{l}{\textbf{QQ RMSE}}\\
\hspace{1em}Conventional & \multicolumn{1}{r}{1.26}\\
\hspace{1em}Distributional 1 & \multicolumn{1}{r}{1.06}\\
\hspace{1em}Distributional 2 & \multicolumn{1}{r}{1.14}\\
\hspace{1em}Distributional 3 & \multicolumn{1}{r}{0.92}\\
\hspace{1em}Distributional 4 & \multicolumn{1}{r}{\textbf{0.62}}\\
\bottomrule
\end{tabular}
\end{table}

\hypertarget{model-specification-details}{%
\subsection{Model specification
details}\label{model-specification-details}}

We modelled the log-ratio dependent variable using the four-parameter
sinh-arcsinh distribution:

\[
\begin{array}{rcl}
y_i &\sim& \sinh(\mu_i, \sigma_i, \epsilon_i, \delta_i) \\
\mu_i &=& \beta^\mu\mathbf{X}^\mu_i \\
\log\sigma^\star_i &=& \beta^\sigma\mathbf{X}^\sigma_i\\
\epsilon_i &=& \beta^\epsilon\mathbf{X}^\epsilon_i \\
\log\delta_i &=& \beta^\delta\mathbf{X}^\delta_i\\
\sigma_i &=& \sigma^\star_i \delta_i,
\end{array}
\]

where \(\beta^\mu\), \(\beta^\sigma\), \(\beta^\epsilon\), and
\(\beta^\delta\) are free parameters. We placed essentially arbitrary
shrinkage priors on all coefficients:

\[
\beta^\mu,\beta^\sigma,\beta^\epsilon,\beta^\delta \sim \text{N}(0, 5).
\]

First, we fit a conventional regression, in which only the location
parameter, \(\mu\), is a function of data. Specifically, we allowed for
linear sex and age effects and a linear interaction between respondent
sex and age (\(s_i\) and \(a_i\), respectively) in the model of \(\mu\):

\[
\begin{array}{rcl}
\mathbf{X}^\mu_i &=& (1 , s_i, a_i, s_i\cdot a_i)\\
\mathbf{X}^\sigma_i, \mathbf{X}^\epsilon_i, \mathbf{X}^\delta_i &=& (1).
\end{array}
\]

In the second model, we allowed the three higher order distributional
parameters to vary by age and sex:

\[
\begin{array}{rcl}
\mathbf{X}^\mu_i &=& (1 , s_i, a_i, s_i\cdot a_i)\\
\mathbf{X}^\sigma_i, \mathbf{X}^\epsilon_i, \mathbf{X}^\delta_i &=& (1, s_i, a_i).
\end{array}
\]

In the third model, all four distributional parameters had age, sex, and
age-sex interaction effects:

\[
\begin{array}{rcl}
\mathbf{X}^\mu_i,\mathbf{X}^\sigma_i, \mathbf{X}^\epsilon_i, \mathbf{X}^\delta_i &=& (1 , s_i, a_i, s_i\cdot a_i)
\end{array}
\]

To allow for the possibility of non-linear variation with respect to age
in the fourth model, we modelled the location parameter using
sex-specific natural splines on age:

\[
\begin{array}{rcl}
\mathbf{X}^\mu_i &=& \left(1 , s_i, \phi_1(a_i), .. .,\phi_K(a_i), s_i\cdot\phi_1(a_i),.. .,s_i\cdot\phi_K(a_i)\right)\\
\mathbf{X}^\sigma_i, \mathbf{X}^\epsilon_i, \mathbf{X}^\delta_i &=& (1, s_i, a_i, s_i\cdot a_i),
\end{array}
\]

where \(K\) is the number of columns in the spline design matrix. By
including a second set of basis function values that are multiplied by
\(s_i\), we are estimating an additional, female-specific trend with
respect to age.

Finally, we fit a fifth model, in which all four distributional
parameters were modelled as sex-specific splines with respect to age:

\[
\begin{array}{rcl}
\mathbf{X}^\mu_i,\mathbf{X}^\sigma_i, \mathbf{X}^\epsilon_i, \mathbf{X}^\delta_i &=& \left(1 , s_i, \phi_1(a_i), .. .,\phi_K(a_i), s_i\cdot\phi_1(a_i),.. .,s_i\cdot\phi_K(a_i)\right).
\end{array}
\]

\hypertarget{full-results}{%
\subsection{Full Results}\label{full-results}}

\hypertarget{supplementary-figures}{%
\subsubsection{Supplementary Figures}\label{supplementary-figures}}
\singlespacing

\begin{figure}[H]
\includegraphics[width=1\linewidth]{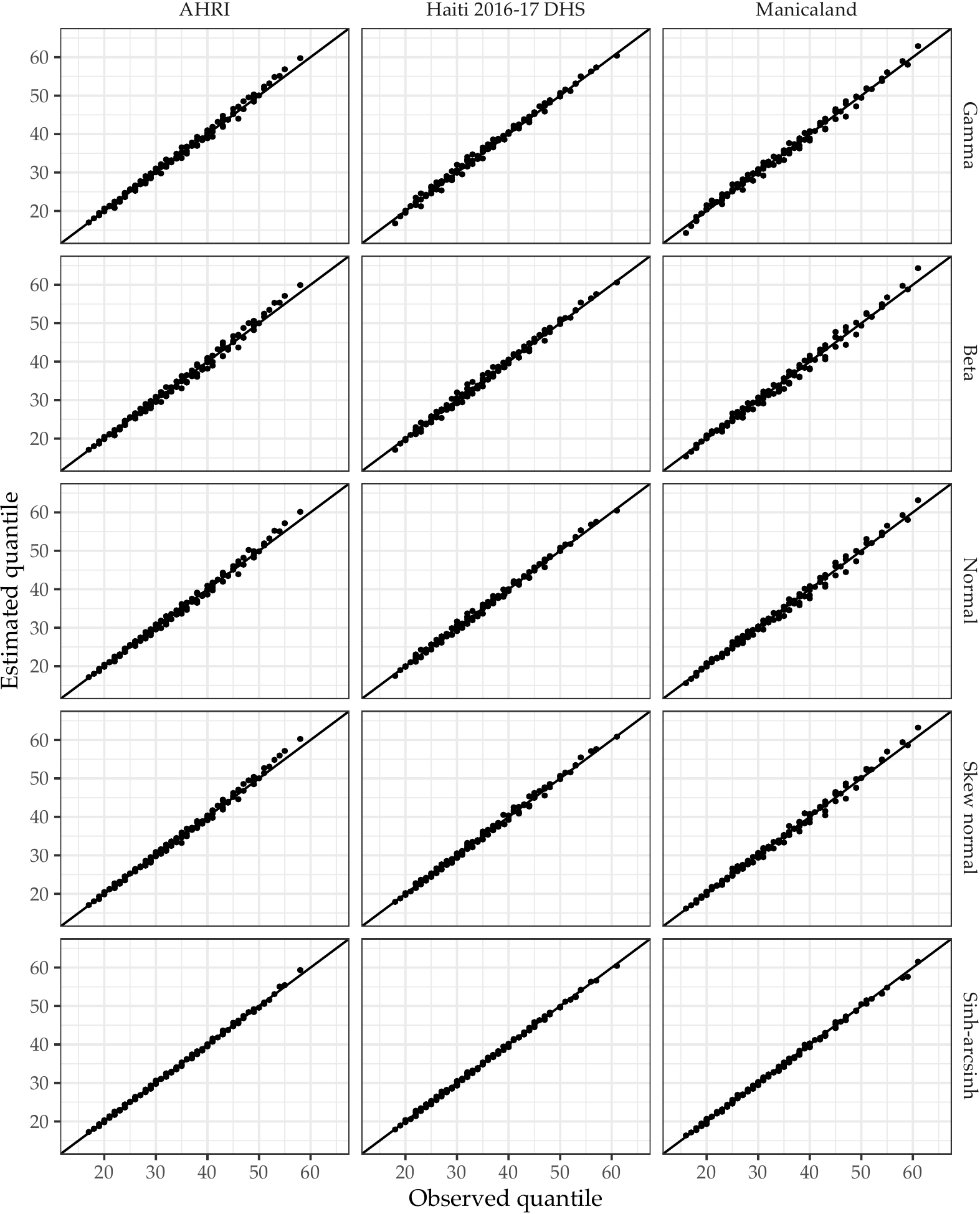} \caption{Overlaid quantile-quantile (QQ) plots for each probability distribution's best fit to data in all three main data sets. Presented quantiles range from 10th to 90th in increments of 10. Lines closer to the line of equality indicate better fit to empirical quantiles}\label{fig:qqPlot}
\end{figure}

\begin{figure}[H]
\includegraphics[width=1\linewidth,]{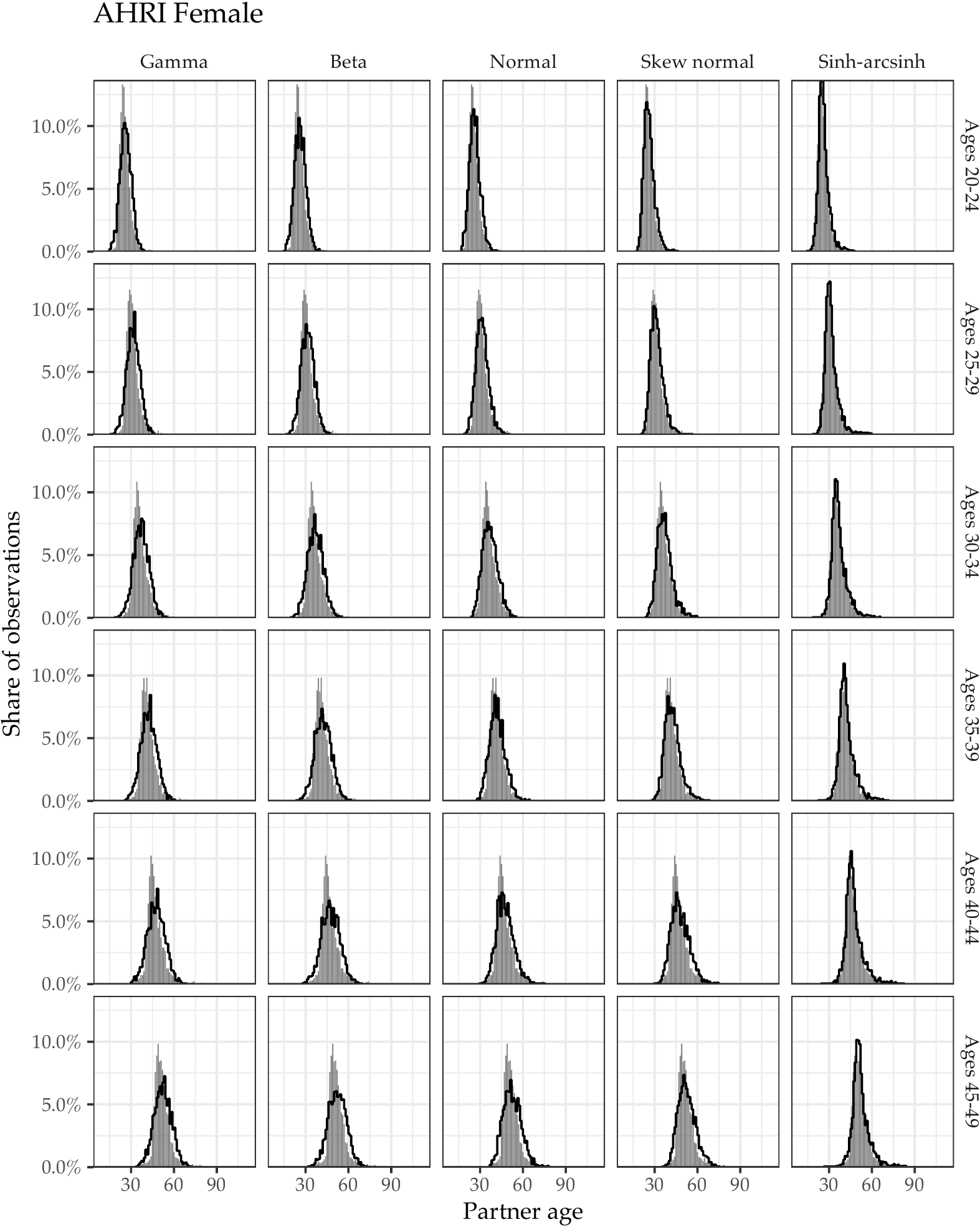} \caption{Observed partner age distributions (grey bars) and posterior predictive partner age distributions (lines) for each probability distribution among women in the AHRI data set. Here, we plot the posterior predicitve distribution associated with each distribution's highest-ELPD dependent variable.}\label{fig:bigHist-1}
\end{figure}
\begin{figure}[H]
\includegraphics[width=1\linewidth,]{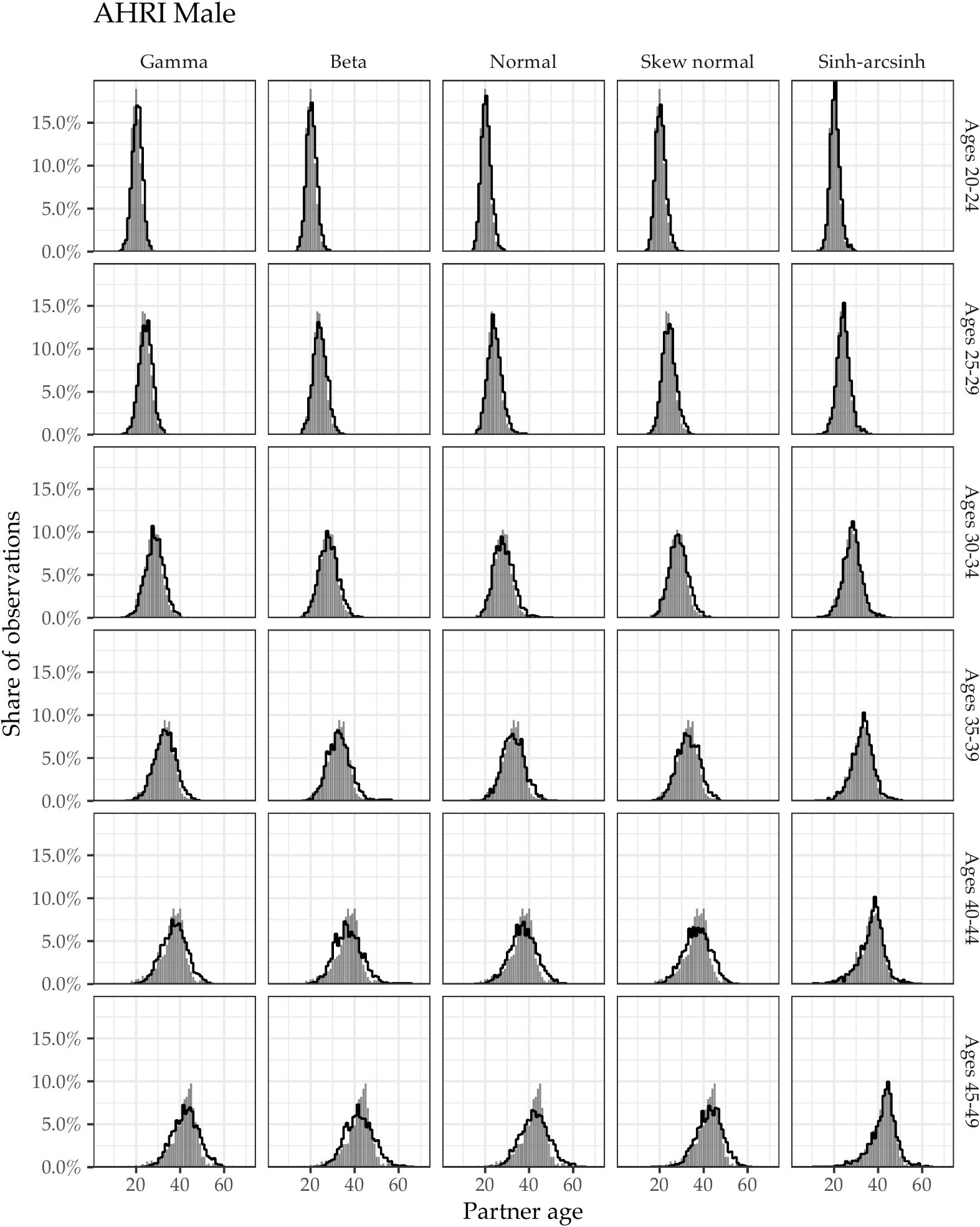} \caption{Observed partner age distributions (grey bars) and posterior predictive partner age distributions (lines) for each probability distribution among men in the AHRI data set. Here, we plot the posterior predicitve distribution associated with each distribution's highest-ELPD dependent variable.}\label{fig:bigHist-2}
\end{figure}
\begin{figure}[H]
\includegraphics[width=1\linewidth,]{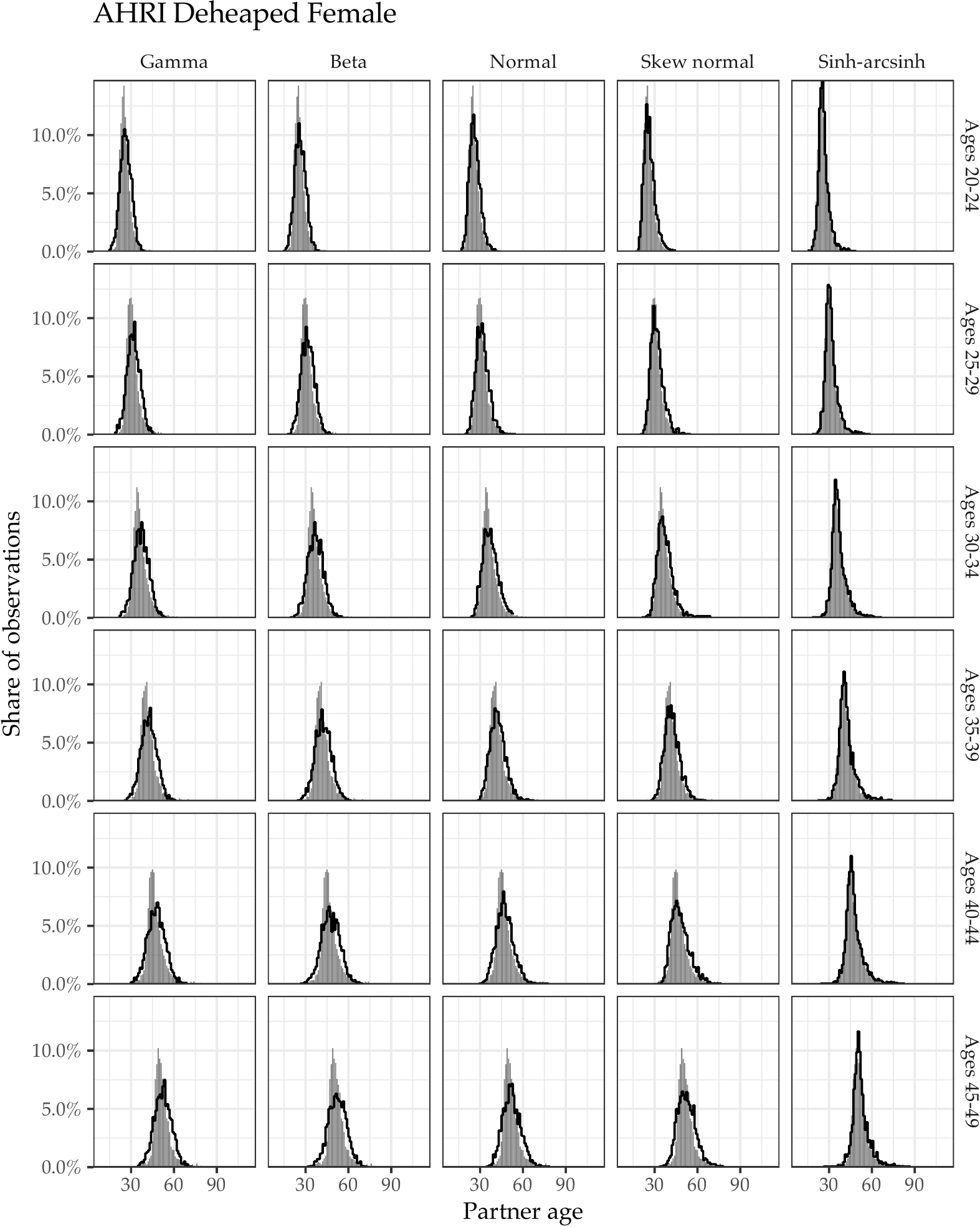} \caption{Observed partner age distributions (grey bars) and posterior predictive partner age distributions (lines) for each probability distribution among women in the AHRI Deheaped data set. Here, we plot the posterior predicitve distribution associated with each distribution's highest-ELPD dependent variable.}\label{fig:bigHist-3}
\end{figure}
\begin{figure}[H]
\includegraphics[width=1\linewidth,]{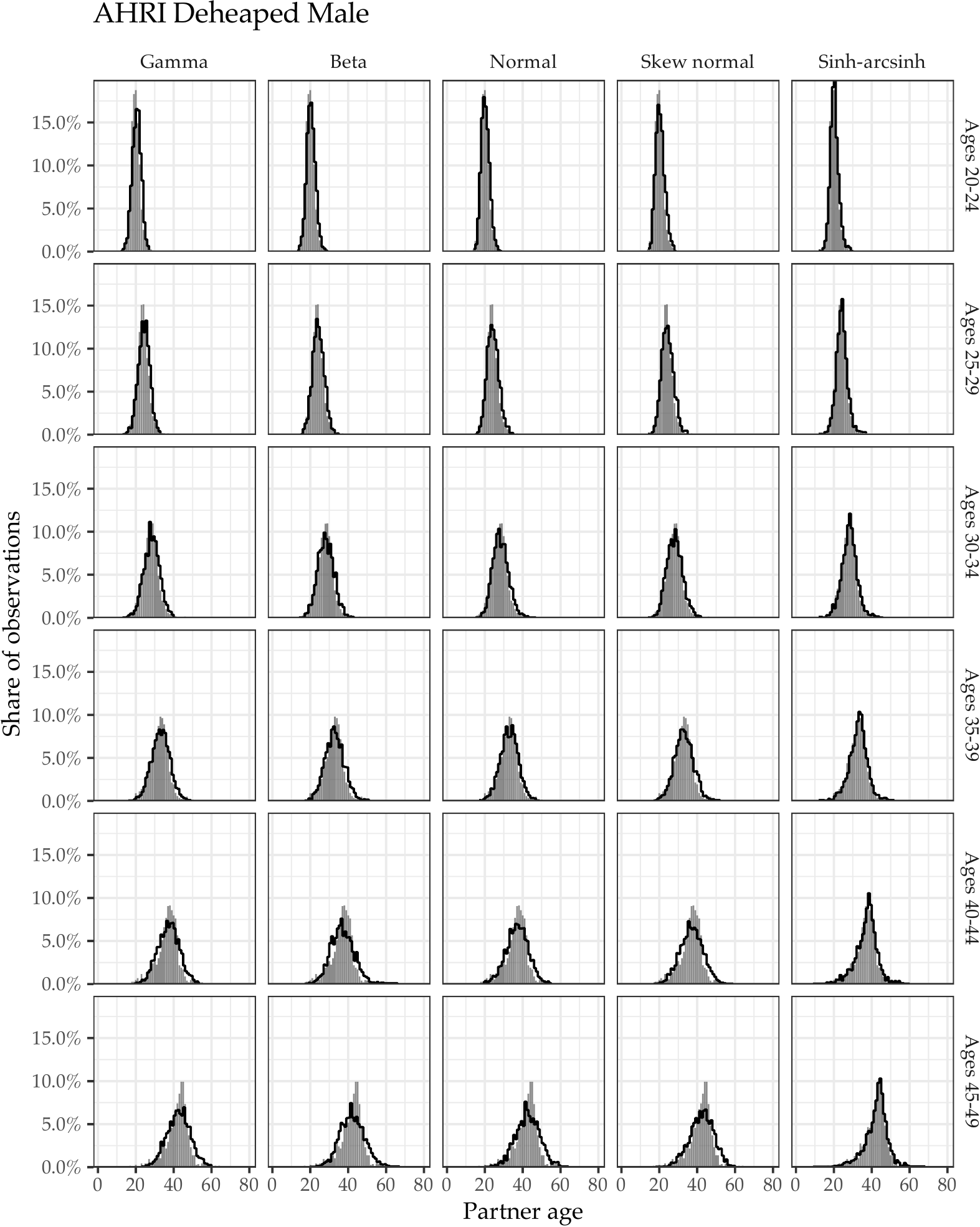} \caption{Observed partner age distributions (grey bars) and posterior predictive partner age distributions (lines) for each probability distribution among men in the AHRI Deheaped data set. Here, we plot the posterior predicitve distribution associated with each distribution's highest-ELPD dependent variable.}\label{fig:bigHist-4}
\end{figure}
\begin{figure}[H]
\includegraphics[width=1\linewidth,]{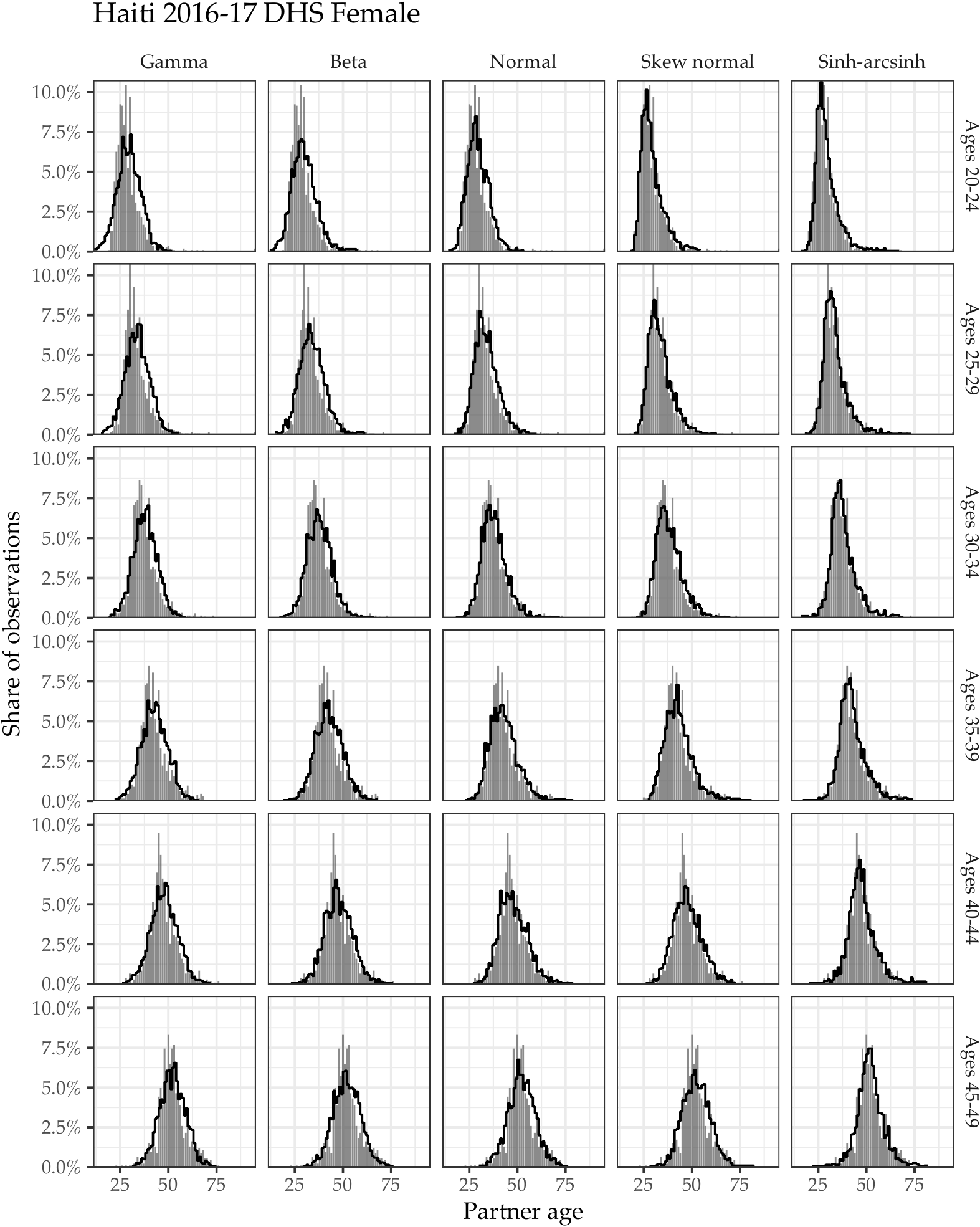} \caption{Observed partner age distributions (grey bars) and posterior predictive partner age distributions (lines) for each probability distribution among women in the Haiti 2016-17 DHS data set. Here, we plot the posterior predicitve distribution associated with each distribution's highest-ELPD dependent variable.}\label{fig:bigHist-5}
\end{figure}
\begin{figure}[H]
\includegraphics[width=1\linewidth,]{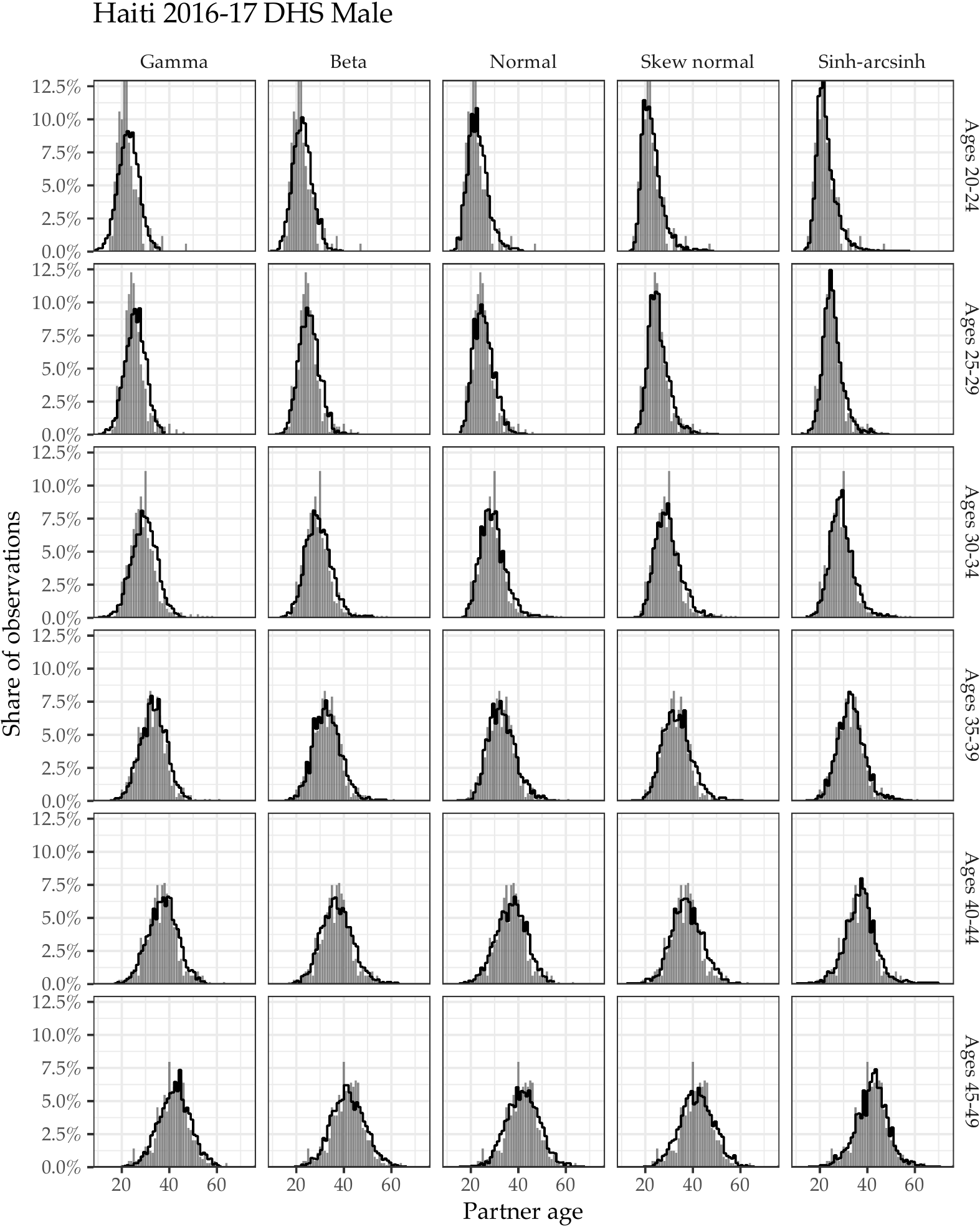} \caption{Observed partner age distributions (grey bars) and posterior predictive partner age distributions (lines) for each probability distribution among men in the Haiti 2016-17 DHS data set. Here, we plot the posterior predicitve distribution associated with each distribution's highest-ELPD dependent variable.}\label{fig:bigHist-6}
\end{figure}
\begin{figure}[H]
\includegraphics[width=1\linewidth,]{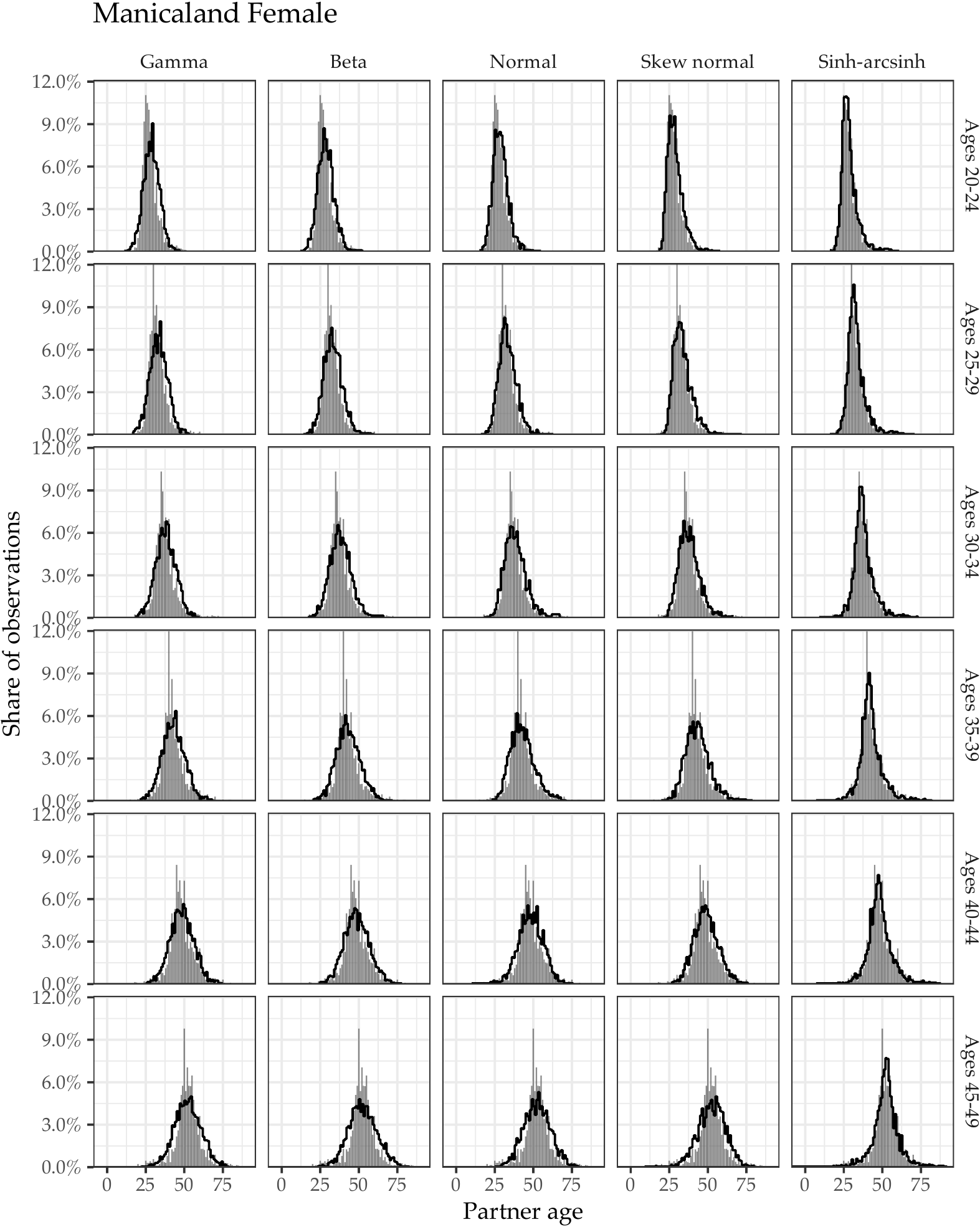} \caption{Observed partner age distributions (grey bars) and posterior predictive partner age distributions (lines) for each probability distribution among women in the Manicaland data set. Here, we plot the posterior predicitve distribution associated with each distribution's highest-ELPD dependent variable.}\label{fig:bigHist-7}
\end{figure}
\begin{figure}[H]
\includegraphics[width=1\linewidth,]{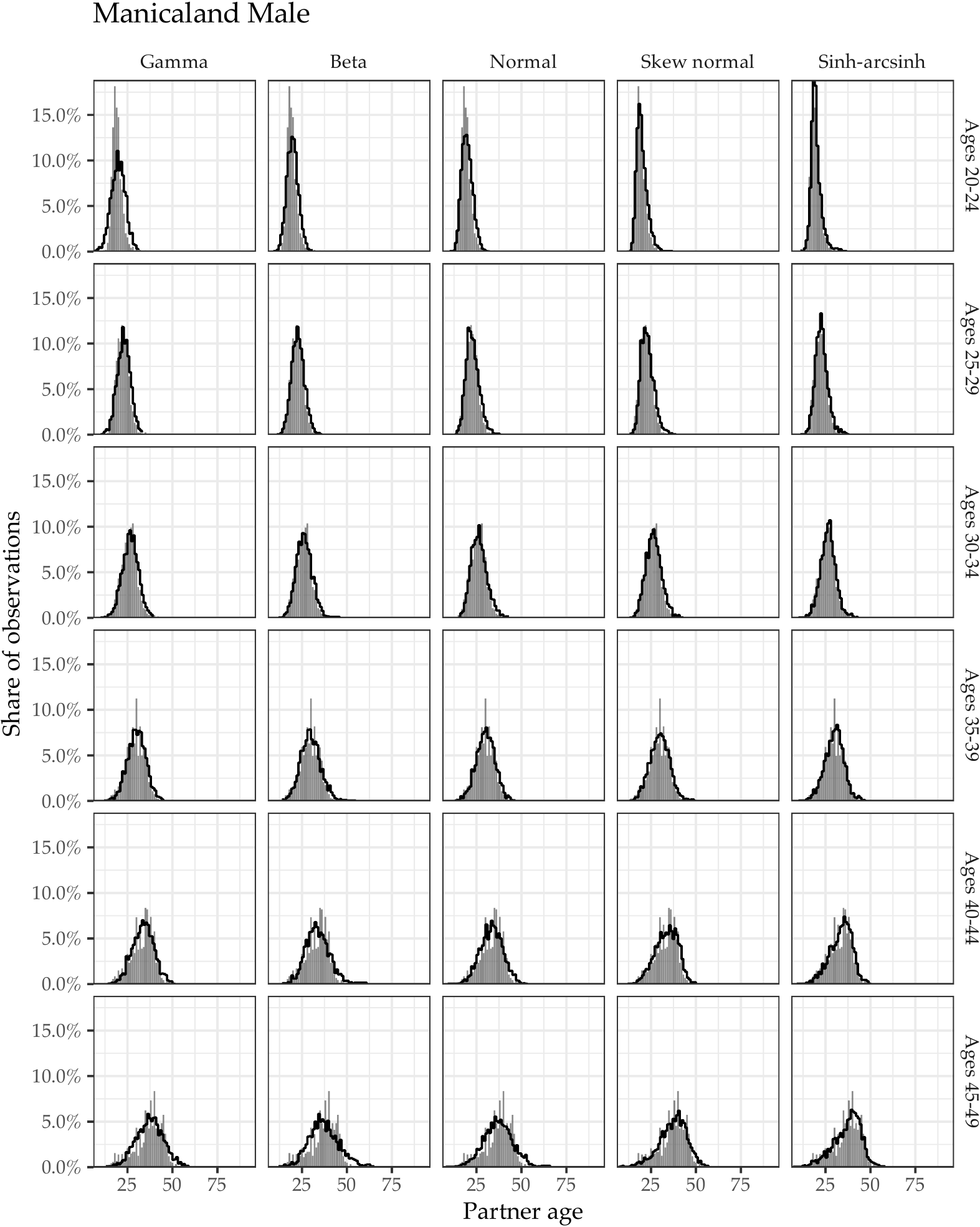} \caption{Observed partner age distributions (grey bars) and posterior predictive partner age distributions (lines) for each probability distribution among men in the Manicaland data set. Here, we plot the posterior predicitve distribution associated with each distribution's highest-ELPD dependent variable.}\label{fig:bigHist-8}
\end{figure}

\newpage

\hypertarget{supplementary-tables}{%
\subsubsection{Supplementary Tables}\label{supplementary-tables}}

\begin{table}[H]

\caption{\label{tab:TableAHRIFemale}Full ELPD and QQ RMSE table for women in the AHRI data set. Higher ELPD values and lower QQ RMSE values are better.}
\centering
\begin{tabular}[t]{rlrrrr}
\toprule
Rank & Distribution & ELPD & ELPD Diff & SE of Diff & QQ RMSE\\
\midrule
\addlinespace[0.3em]
\multicolumn{6}{l}{\textbf{AHRI Female 20-24}}\\
\hspace{1em}1 & Sinh-arcsinh & -31750.94 & 0.00 & 0.00 & 0.32\\
\hspace{1em}2 & Skew normal & -32056.39 & -305.46 & 48.63 & 0.47\\
\hspace{1em}3 & Normal & -32414.54 & -663.61 & 60.54 & 0.62\\
\hspace{1em}4 & Beta & -32953.92 & -1202.98 & 112.08 & 0.77\\
\hspace{1em}5 & Gamma & -33461.85 & -1710.92 & 148.15 & 0.80\\
\addlinespace[0.3em]
\multicolumn{6}{l}{\textbf{AHRI Female 25-29}}\\
\hspace{1em}1 & Sinh-arcsinh & -24647.65 & 0.00 & 0.00 & 0.28\\
\hspace{1em}2 & Skew normal & -24906.22 & -258.57 & 43.27 & 0.52\\
\hspace{1em}3 & Normal & -25238.71 & -591.06 & 54.82 & 0.68\\
\hspace{1em}4 & Beta & -25701.13 & -1053.48 & 114.84 & 0.89\\
\hspace{1em}5 & Gamma & -25995.81 & -1348.16 & 132.15 & 0.90\\
\addlinespace[0.3em]
\multicolumn{6}{l}{\textbf{AHRI Female 30-34}}\\
\hspace{1em}1 & Sinh-arcsinh & -19831.53 & 0.00 & 0.00 & 0.44\\
\hspace{1em}2 & Skew normal & -20200.44 & -368.91 & 69.40 & 0.51\\
\hspace{1em}3 & Normal & -20314.79 & -483.26 & 52.24 & 0.80\\
\hspace{1em}4 & Beta & -20575.61 & -744.08 & 67.46 & 0.93\\
\hspace{1em}5 & Gamma & -20708.35 & -876.82 & 73.89 & 0.91\\
\addlinespace[0.3em]
\multicolumn{6}{l}{\textbf{AHRI Female 35-39}}\\
\hspace{1em}1 & Sinh-arcsinh & -15469.18 & 0.00 & 0.00 & 0.31\\
\hspace{1em}2 & Skew normal & -15749.79 & -280.61 & 53.04 & 0.77\\
\hspace{1em}3 & Normal & -15834.32 & -365.14 & 41.23 & 0.80\\
\hspace{1em}4 & Beta & -16026.51 & -557.33 & 53.99 & 1.18\\
\hspace{1em}5 & Gamma & -16087.40 & -618.22 & 57.06 & 1.04\\
\addlinespace[0.3em]
\multicolumn{6}{l}{\textbf{AHRI Female 40-44}}\\
\hspace{1em}1 & Sinh-arcsinh & -12556.61 & 0.00 & 0.00 & 0.45\\
\hspace{1em}2 & Skew normal & -12876.71 & -320.10 & 45.85 & 1.27\\
\hspace{1em}3 & Normal & -12935.34 & -378.73 & 52.38 & 0.92\\
\hspace{1em}4 & Beta & -13137.69 & -581.08 & 69.18 & 1.38\\
\hspace{1em}5 & Gamma & -13150.66 & -594.05 & 62.73 & 1.19\\
\addlinespace[0.3em]
\multicolumn{6}{l}{\textbf{AHRI Female 45-49}}\\
\hspace{1em}1 & Sinh-arcsinh & -10059.21 & 0.00 & 0.00 & 0.59\\
\hspace{1em}2 & Skew normal & -10391.95 & -332.74 & 42.75 & 1.36\\
\hspace{1em}3 & Normal & -10433.64 & -374.43 & 48.91 & 1.53\\
\hspace{1em}4 & Gamma & -10527.00 & -467.79 & 50.72 & 1.35\\
\hspace{1em}5 & Beta & -10545.33 & -486.12 & 56.02 & 1.58\\
\bottomrule
\end{tabular}
\end{table}
\begin{table}[H]

\caption{\label{tab:TableAHRIMale}Full ELPD and QQ RMSE table for men in the AHRI data set. Higher ELPD values and lower QQ RMSE values are better.}
\centering
\begin{tabular}[t]{rlrrrr}
\toprule
Rank & Distribution & ELPD & ELPD Diff & SE of Diff & QQ RMSE\\
\midrule
\addlinespace[0.3em]
\multicolumn{6}{l}{\textbf{AHRI Male 20-24}}\\
\hspace{1em}1 & Sinh-arcsinh & -20428.11 & 0.00 & 0.00 & 0.23\\
\hspace{1em}2 & Skew normal & -20499.86 & -71.75 & 17.12 & 0.25\\
\hspace{1em}3 & Normal & -20503.89 & -75.79 & 16.85 & 0.22\\
\hspace{1em}4 & Beta & -20545.59 & -117.49 & 23.21 & 0.22\\
\hspace{1em}5 & Gamma & -20700.24 & -272.13 & 43.53 & 0.29\\
\addlinespace[0.3em]
\multicolumn{6}{l}{\textbf{AHRI Male 25-29}}\\
\hspace{1em}1 & Sinh-arcsinh & -12664.21 & 0.00 & 0.00 & 0.26\\
\hspace{1em}2 & Skew normal & -12727.03 & -62.82 & 17.86 & 0.28\\
\hspace{1em}3 & Beta & -12739.03 & -74.82 & 18.65 & 0.31\\
\hspace{1em}4 & Normal & -12753.25 & -89.04 & 19.35 & 0.29\\
\hspace{1em}5 & Gamma & -12788.26 & -124.05 & 35.07 & 0.38\\
\addlinespace[0.3em]
\multicolumn{6}{l}{\textbf{AHRI Male 30-34}}\\
\hspace{1em}1 & Sinh-arcsinh & -9301.03 & 0.00 & 0.00 & 0.29\\
\hspace{1em}2 & Skew normal & -9357.18 & -56.15 & 14.08 & 0.43\\
\hspace{1em}3 & Beta & -9371.86 & -70.83 & 16.48 & 0.37\\
\hspace{1em}4 & Normal & -9385.63 & -84.60 & 14.67 & 0.46\\
\hspace{1em}5 & Gamma & -9419.34 & -118.31 & 35.11 & 0.27\\
\addlinespace[0.3em]
\multicolumn{6}{l}{\textbf{AHRI Male 35-39}}\\
\hspace{1em}1 & Sinh-arcsinh & -6746.89 & 0.00 & 0.00 & 0.30\\
\hspace{1em}2 & Skew normal & -6812.77 & -65.88 & 17.73 & 0.64\\
\hspace{1em}3 & Normal & -6817.86 & -70.97 & 23.24 & 0.70\\
\hspace{1em}4 & Beta & -6830.95 & -84.06 & 17.95 & 0.71\\
\hspace{1em}5 & Gamma & -6832.47 & -85.58 & 32.03 & 0.44\\
\addlinespace[0.3em]
\multicolumn{6}{l}{\textbf{AHRI Male 40-44}}\\
\hspace{1em}1 & Sinh-arcsinh & -4610.95 & 0.00 & 0.00 & 0.35\\
\hspace{1em}2 & Skew normal & -4711.78 & -100.84 & 18.66 & 0.92\\
\hspace{1em}3 & Normal & -4713.78 & -102.83 & 18.82 & 0.78\\
\hspace{1em}4 & Gamma & -4718.28 & -107.33 & 24.83 & 0.63\\
\hspace{1em}5 & Beta & -4742.70 & -131.75 & 17.28 & 1.07\\
\addlinespace[0.3em]
\multicolumn{6}{l}{\textbf{AHRI Male 45-49}}\\
\hspace{1em}1 & Sinh-arcsinh & -3683.47 & 0.00 & 0.00 & 0.34\\
\hspace{1em}2 & Skew normal & -3770.59 & -87.12 & 16.56 & 0.81\\
\hspace{1em}3 & Gamma & -3776.33 & -92.86 & 15.56 & 0.87\\
\hspace{1em}4 & Normal & -3778.84 & -95.37 & 14.50 & 1.17\\
\hspace{1em}5 & Beta & -3805.78 & -122.31 & 17.40 & 1.36\\
\bottomrule
\end{tabular}
\end{table}
\begin{table}[H]

\caption{\label{tab:TableAHRI DeheapedFemale}Full ELPD and QQ RMSE table for women in the AHRI Deheaped data set. Higher ELPD values and lower QQ RMSE values are better.}
\centering
\begin{tabular}[t]{rlrrrr}
\toprule
Rank & Distribution & ELPD & ELPD Diff & SE of Diff & QQ RMSE\\
\midrule
\addlinespace[0.3em]
\multicolumn{6}{l}{\textbf{AHRI Deheaped Female 20-24}}\\
\hspace{1em}1 & Sinh-arcsinh & -31411.24 & 0.00 & 0.00 & 0.26\\
\hspace{1em}2 & Skew normal & -31797.37 & -386.13 & 53.15 & 0.59\\
\hspace{1em}3 & Normal & -32179.29 & -768.05 & 65.50 & 0.56\\
\hspace{1em}4 & Beta & -32737.57 & -1326.32 & 118.47 & 0.76\\
\hspace{1em}5 & Gamma & -33254.17 & -1842.92 & 155.14 & 0.78\\
\addlinespace[0.3em]
\multicolumn{6}{l}{\textbf{AHRI Deheaped Female 25-29}}\\
\hspace{1em}1 & Sinh-arcsinh & -24439.47 & 0.00 & 0.00 & 0.27\\
\hspace{1em}2 & Skew normal & -24768.06 & -328.59 & 46.71 & 0.65\\
\hspace{1em}3 & Normal & -25104.46 & -664.99 & 58.32 & 0.82\\
\hspace{1em}4 & Beta & -25574.33 & -1134.86 & 119.65 & 1.03\\
\hspace{1em}5 & Gamma & -25870.30 & -1430.83 & 137.51 & 1.05\\
\addlinespace[0.3em]
\multicolumn{6}{l}{\textbf{AHRI Deheaped Female 30-34}}\\
\hspace{1em}1 & Sinh-arcsinh & -19680.77 & 0.00 & 0.00 & 0.41\\
\hspace{1em}2 & Skew normal & -20112.70 & -431.94 & 72.95 & 0.55\\
\hspace{1em}3 & Normal & -20228.52 & -547.76 & 56.19 & 0.81\\
\hspace{1em}4 & Beta & -20492.23 & -811.46 & 70.53 & 0.92\\
\hspace{1em}5 & Gamma & -20624.82 & -944.06 & 76.98 & 0.80\\
\addlinespace[0.3em]
\multicolumn{6}{l}{\textbf{AHRI Deheaped Female 35-39}}\\
\hspace{1em}1 & Sinh-arcsinh & -15381.68 & 0.00 & 0.00 & 0.26\\
\hspace{1em}2 & Skew normal & -15703.77 & -322.09 & 55.30 & 0.68\\
\hspace{1em}3 & Normal & -15788.73 & -407.05 & 43.67 & 0.82\\
\hspace{1em}4 & Beta & -15983.57 & -601.90 & 56.31 & 1.13\\
\hspace{1em}5 & Gamma & -16044.22 & -662.54 & 59.17 & 1.04\\
\addlinespace[0.3em]
\multicolumn{6}{l}{\textbf{AHRI Deheaped Female 40-44}}\\
\hspace{1em}1 & Sinh-arcsinh & -12491.91 & 0.00 & 0.00 & 0.25\\
\hspace{1em}2 & Skew normal & -12846.63 & -354.72 & 47.38 & 0.99\\
\hspace{1em}3 & Normal & -12905.04 & -413.12 & 54.14 & 0.89\\
\hspace{1em}4 & Beta & -13109.82 & -617.91 & 70.96 & 1.31\\
\hspace{1em}5 & Gamma & -13121.45 & -629.53 & 64.12 & 1.13\\
\addlinespace[0.3em]
\multicolumn{6}{l}{\textbf{AHRI Deheaped Female 45-49}}\\
\hspace{1em}1 & Sinh-arcsinh & -9981.83 & 0.00 & 0.00 & 0.53\\
\hspace{1em}2 & Skew normal & -10357.85 & -376.01 & 45.08 & 1.43\\
\hspace{1em}3 & Normal & -10401.64 & -419.80 & 51.57 & 1.46\\
\hspace{1em}4 & Gamma & -10493.73 & -511.90 & 52.90 & 1.37\\
\hspace{1em}5 & Beta & -10513.46 & -531.63 & 58.21 & 1.61\\
\bottomrule
\end{tabular}
\end{table}
\begin{table}[H]

\caption{\label{tab:TableAHRI DeheapedMale}Full ELPD and QQ RMSE table for men in the AHRI Deheaped data set. Higher ELPD values and lower QQ RMSE values are better.}
\centering
\begin{tabular}[t]{rlrrrr}
\toprule
Rank & Distribution & ELPD & ELPD Diff & SE of Diff & QQ RMSE\\
\midrule
\addlinespace[0.3em]
\multicolumn{6}{l}{\textbf{AHRI Deheaped Male 20-24}}\\
\hspace{1em}1 & Sinh-arcsinh & -20310.35 & 0.00 & 0.00 & 0.27\\
\hspace{1em}2 & Skew normal & -20429.90 & -119.55 & 27.09 & 0.22\\
\hspace{1em}3 & Normal & -20459.73 & -149.38 & 35.78 & 0.29\\
\hspace{1em}4 & Beta & -20574.15 & -263.80 & 75.13 & 0.22\\
\hspace{1em}5 & Gamma & -20899.52 & -589.17 & 175.99 & 0.27\\
\addlinespace[0.3em]
\multicolumn{6}{l}{\textbf{AHRI Deheaped Male 25-29}}\\
\hspace{1em}1 & Sinh-arcsinh & -12585.54 & 0.00 & 0.00 & 0.28\\
\hspace{1em}2 & Skew normal & -12680.59 & -95.05 & 21.53 & 0.44\\
\hspace{1em}3 & Beta & -12697.00 & -111.46 & 23.31 & 0.37\\
\hspace{1em}4 & Normal & -12701.76 & -116.23 & 22.96 & 0.41\\
\hspace{1em}5 & Gamma & -12763.81 & -178.27 & 41.24 & 0.39\\
\addlinespace[0.3em]
\multicolumn{6}{l}{\textbf{AHRI Deheaped Male 30-34}}\\
\hspace{1em}1 & Sinh-arcsinh & -9227.26 & 0.00 & 0.00 & 0.37\\
\hspace{1em}2 & Skew normal & -9302.42 & -75.16 & 16.15 & 0.41\\
\hspace{1em}3 & Beta & -9318.24 & -90.97 & 19.07 & 0.39\\
\hspace{1em}4 & Normal & -9327.58 & -100.31 & 16.18 & 0.41\\
\hspace{1em}5 & Gamma & -9372.32 & -145.06 & 38.27 & 0.27\\
\addlinespace[0.3em]
\multicolumn{6}{l}{\textbf{AHRI Deheaped Male 35-39}}\\
\hspace{1em}1 & Sinh-arcsinh & -6694.86 & 0.00 & 0.00 & 0.30\\
\hspace{1em}2 & Skew normal & -6774.11 & -79.26 & 19.32 & 0.61\\
\hspace{1em}3 & Normal & -6780.69 & -85.84 & 25.42 & 0.44\\
\hspace{1em}4 & Beta & -6791.95 & -97.10 & 19.81 & 0.69\\
\hspace{1em}5 & Gamma & -6796.41 & -101.55 & 34.45 & 0.40\\
\addlinespace[0.3em]
\multicolumn{6}{l}{\textbf{AHRI Deheaped Male 40-44}}\\
\hspace{1em}1 & Sinh-arcsinh & -4591.04 & 0.00 & 0.00 & 0.49\\
\hspace{1em}2 & Skew normal & -4700.54 & -109.51 & 19.38 & 1.16\\
\hspace{1em}3 & Normal & -4703.52 & -112.49 & 19.93 & 1.00\\
\hspace{1em}4 & Gamma & -4708.43 & -117.40 & 25.94 & 0.89\\
\hspace{1em}5 & Beta & -4731.41 & -140.37 & 17.84 & 1.30\\
\addlinespace[0.3em]
\multicolumn{6}{l}{\textbf{AHRI Deheaped Male 45-49}}\\
\hspace{1em}1 & Sinh-arcsinh & -3680.18 & 0.00 & 0.00 & 0.30\\
\hspace{1em}2 & Normal & -3796.06 & -115.88 & 19.24 & 1.15\\
\hspace{1em}3 & Skew normal & -3797.14 & -116.95 & 23.48 & 1.02\\
\hspace{1em}4 & Gamma & -3801.02 & -120.83 & 24.51 & 0.98\\
\hspace{1em}5 & Beta & -3817.97 & -137.79 & 19.37 & 1.39\\
\bottomrule
\end{tabular}
\end{table}
\begin{table}[H]

\caption{\label{tab:TableHaiti 2016-17 DHSFemale}Full ELPD and QQ RMSE table for women in the Haiti 2016-17 DHS data set. Higher ELPD values and lower QQ RMSE values are better.}
\centering
\begin{tabular}[t]{rlrrrr}
\toprule
Rank & Distribution & ELPD & ELPD Diff & SE of Diff & QQ RMSE\\
\midrule
\addlinespace[0.3em]
\multicolumn{6}{l}{\textbf{Haiti 2016-17 DHS Female 20-24}}\\
\hspace{1em}1 & Sinh-arcsinh & -3259.31 & 0.00 & 0.00 & 0.49\\
\hspace{1em}2 & Skew normal & -3263.46 & -4.15 & 4.95 & 0.53\\
\hspace{1em}3 & Normal & -3338.23 & -78.92 & 19.54 & 0.91\\
\hspace{1em}4 & Beta & -3441.91 & -182.60 & 45.77 & 1.24\\
\hspace{1em}5 & Gamma & -3504.85 & -245.54 & 53.90 & 1.29\\
\addlinespace[0.3em]
\multicolumn{6}{l}{\textbf{Haiti 2016-17 DHS Female 25-29}}\\
\hspace{1em}1 & Sinh-arcsinh & -4447.43 & 0.00 & 0.00 & 0.26\\
\hspace{1em}2 & Skew normal & -4471.22 & -23.78 & 8.41 & 0.57\\
\hspace{1em}3 & Normal & -4527.25 & -79.82 & 18.72 & 0.86\\
\hspace{1em}4 & Beta & -4625.97 & -178.54 & 40.88 & 1.23\\
\hspace{1em}5 & Gamma & -4678.20 & -230.77 & 45.81 & 1.22\\
\addlinespace[0.3em]
\multicolumn{6}{l}{\textbf{Haiti 2016-17 DHS Female 30-34}}\\
\hspace{1em}1 & Sinh-arcsinh & -4720.12 & 0.00 & 0.00 & 0.44\\
\hspace{1em}2 & Skew normal & -4749.57 & -29.45 & 9.06 & 0.68\\
\hspace{1em}3 & Normal & -4763.78 & -43.66 & 10.51 & 0.62\\
\hspace{1em}4 & Beta & -4809.11 & -88.99 & 17.19 & 0.85\\
\hspace{1em}5 & Gamma & -4836.82 & -116.70 & 20.32 & 0.83\\
\addlinespace[0.3em]
\multicolumn{6}{l}{\textbf{Haiti 2016-17 DHS Female 35-39}}\\
\hspace{1em}1 & Sinh-arcsinh & -4490.82 & 0.00 & 0.00 & 0.33\\
\hspace{1em}2 & Skew normal & -4518.58 & -27.75 & 8.14 & 0.57\\
\hspace{1em}3 & Normal & -4526.55 & -35.73 & 8.59 & 0.73\\
\hspace{1em}4 & Beta & -4561.27 & -70.45 & 13.60 & 0.94\\
\hspace{1em}5 & Gamma & -4577.84 & -87.01 & 15.27 & 0.86\\
\addlinespace[0.3em]
\multicolumn{6}{l}{\textbf{Haiti 2016-17 DHS Female 40-44}}\\
\hspace{1em}1 & Sinh-arcsinh & -3601.02 & 0.00 & 0.00 & 0.35\\
\hspace{1em}2 & Skew normal & -3629.45 & -28.43 & 7.51 & 0.83\\
\hspace{1em}3 & Normal & -3633.14 & -32.11 & 7.96 & 0.71\\
\hspace{1em}4 & Beta & -3641.61 & -40.59 & 9.76 & 0.73\\
\hspace{1em}5 & Gamma & -3644.89 & -43.86 & 10.47 & 0.64\\
\addlinespace[0.3em]
\multicolumn{6}{l}{\textbf{Haiti 2016-17 DHS Female 45-49}}\\
\hspace{1em}1 & Sinh-arcsinh & -3106.27 & 0.00 & 0.00 & 0.39\\
\hspace{1em}2 & Skew normal & -3133.10 & -26.82 & 7.68 & 0.88\\
\hspace{1em}3 & Gamma & -3133.61 & -27.33 & 7.50 & 0.68\\
\hspace{1em}4 & Normal & -3134.62 & -28.35 & 7.46 & 0.81\\
\hspace{1em}5 & Beta & -3136.89 & -30.62 & 8.61 & 0.88\\
\bottomrule
\end{tabular}
\end{table}
\begin{table}[H]

\caption{\label{tab:TableHaiti 2016-17 DHSMale}Full ELPD and QQ RMSE table for men in the Haiti 2016-17 DHS data set. Higher ELPD values and lower QQ RMSE values are better.}
\centering
\begin{tabular}[t]{rlrrrr}
\toprule
Rank & Distribution & ELPD & ELPD Diff & SE of Diff & QQ RMSE\\
\midrule
\addlinespace[0.3em]
\multicolumn{6}{l}{\textbf{Haiti 2016-17 DHS Male 20-24}}\\
\hspace{1em}1 & Skew normal & -468.98 & 0.00 & 0.00 & 0.43\\
\hspace{1em}2 & Sinh-arcsinh & -469.60 & -0.62 & 1.12 & 0.41\\
\hspace{1em}3 & Normal & -475.31 & -6.33 & 4.28 & 0.67\\
\hspace{1em}4 & Beta & -483.53 & -14.55 & 7.33 & 0.65\\
\hspace{1em}5 & Gamma & -500.53 & -31.55 & 13.35 & 0.94\\
\addlinespace[0.3em]
\multicolumn{6}{l}{\textbf{Haiti 2016-17 DHS Male 25-29}}\\
\hspace{1em}1 & Sinh-arcsinh & -1386.13 & 0.00 & 0.00 & 0.38\\
\hspace{1em}2 & Skew normal & -1390.54 & -4.41 & 3.19 & 0.49\\
\hspace{1em}3 & Normal & -1395.47 & -9.34 & 4.79 & 0.60\\
\hspace{1em}4 & Beta & -1407.46 & -21.32 & 7.31 & 0.62\\
\hspace{1em}5 & Gamma & -1434.18 & -48.04 & 11.75 & 0.79\\
\addlinespace[0.3em]
\multicolumn{6}{l}{\textbf{Haiti 2016-17 DHS Male 30-34}}\\
\hspace{1em}1 & Sinh-arcsinh & -2217.20 & 0.00 & 0.00 & 0.44\\
\hspace{1em}2 & Skew normal & -2222.10 & -4.89 & 3.42 & 0.69\\
\hspace{1em}3 & Normal & -2223.97 & -6.76 & 4.48 & 0.45\\
\hspace{1em}4 & Beta & -2240.58 & -23.37 & 9.32 & 0.52\\
\hspace{1em}5 & Gamma & -2281.18 & -63.98 & 17.91 & 0.73\\
\addlinespace[0.3em]
\multicolumn{6}{l}{\textbf{Haiti 2016-17 DHS Male 35-39}}\\
\hspace{1em}1 & Sinh-arcsinh & -2185.96 & 0.00 & 0.00 & 0.28\\
\hspace{1em}2 & Skew normal & -2189.87 & -3.91 & 2.67 & 0.69\\
\hspace{1em}3 & Beta & -2191.05 & -5.10 & 3.68 & 0.48\\
\hspace{1em}4 & Normal & -2191.11 & -5.16 & 3.55 & 0.49\\
\hspace{1em}5 & Gamma & -2205.69 & -19.73 & 9.57 & 0.52\\
\addlinespace[0.3em]
\multicolumn{6}{l}{\textbf{Haiti 2016-17 DHS Male 40-44}}\\
\hspace{1em}1 & Sinh-arcsinh & -2051.62 & 0.00 & 0.00 & 0.39\\
\hspace{1em}2 & Skew normal & -2060.16 & -8.54 & 4.21 & 0.72\\
\hspace{1em}3 & Normal & -2060.38 & -8.75 & 4.57 & 0.69\\
\hspace{1em}4 & Beta & -2062.00 & -10.37 & 4.87 & 0.70\\
\hspace{1em}5 & Gamma & -2063.79 & -12.17 & 5.73 & 0.47\\
\addlinespace[0.3em]
\multicolumn{6}{l}{\textbf{Haiti 2016-17 DHS Male 45-49}}\\
\hspace{1em}1 & Sinh-arcsinh & -2138.34 & 0.00 & 0.00 & 0.23\\
\hspace{1em}2 & Normal & -2150.53 & -12.19 & 6.38 & 0.35\\
\hspace{1em}3 & Skew normal & -2151.51 & -13.17 & 5.97 & 0.56\\
\hspace{1em}4 & Gamma & -2152.88 & -14.54 & 8.93 & 0.25\\
\hspace{1em}5 & Beta & -2156.13 & -17.79 & 6.14 & 0.48\\
\bottomrule
\end{tabular}
\end{table}
\begin{table}[H]

\caption{\label{tab:TableManicalandFemale}Full ELPD and QQ RMSE table for women in the Manicaland data set. Higher ELPD values and lower QQ RMSE values are better.}
\centering
\begin{tabular}[t]{rlrrrr}
\toprule
Rank & Distribution & ELPD & ELPD Diff & SE of Diff & QQ RMSE\\
\midrule
\addlinespace[0.3em]
\multicolumn{6}{l}{\textbf{Manicaland Female 20-24}}\\
\hspace{1em}1 & Sinh-arcsinh & -16390.77 & 0.00 & 0.00 & 0.31\\
\hspace{1em}2 & Skew normal & -16502.01 & -111.25 & 21.22 & 0.44\\
\hspace{1em}3 & Normal & -16779.93 & -389.16 & 37.05 & 0.67\\
\hspace{1em}4 & Beta & -17111.57 & -720.80 & 62.02 & 0.86\\
\hspace{1em}5 & Gamma & -17387.38 & -996.61 & 76.80 & 1.02\\
\addlinespace[0.3em]
\multicolumn{6}{l}{\textbf{Manicaland Female 25-29}}\\
\hspace{1em}1 & Sinh-arcsinh & -18702.50 & 0.00 & 0.00 & 0.53\\
\hspace{1em}2 & Skew normal & -18923.04 & -220.53 & 25.27 & 0.94\\
\hspace{1em}3 & Normal & -19080.66 & -378.16 & 36.05 & 0.83\\
\hspace{1em}4 & Beta & -19405.80 & -703.30 & 64.97 & 1.05\\
\hspace{1em}5 & Gamma & -19615.53 & -913.03 & 76.38 & 1.09\\
\addlinespace[0.3em]
\multicolumn{6}{l}{\textbf{Manicaland Female 30-34}}\\
\hspace{1em}1 & Sinh-arcsinh & -16523.81 & 0.00 & 0.00 & 0.48\\
\hspace{1em}2 & Skew normal & -16877.96 & -354.15 & 40.36 & 0.87\\
\hspace{1em}3 & Normal & -16886.62 & -362.80 & 36.41 & 0.99\\
\hspace{1em}4 & Beta & -17021.26 & -497.44 & 43.60 & 1.12\\
\hspace{1em}5 & Gamma & -17094.58 & -570.76 & 49.53 & 0.93\\
\addlinespace[0.3em]
\multicolumn{6}{l}{\textbf{Manicaland Female 35-39}}\\
\hspace{1em}1 & Sinh-arcsinh & -14397.76 & 0.00 & 0.00 & 0.48\\
\hspace{1em}2 & Skew normal & -14736.64 & -338.88 & 28.35 & 1.25\\
\hspace{1em}3 & Normal & -14798.55 & -400.79 & 36.87 & 1.39\\
\hspace{1em}4 & Beta & -14824.80 & -427.04 & 33.02 & 1.47\\
\hspace{1em}5 & Gamma & -14835.11 & -437.35 & 34.49 & 1.14\\
\addlinespace[0.3em]
\multicolumn{6}{l}{\textbf{Manicaland Female 40-44}}\\
\hspace{1em}1 & Sinh-arcsinh & -12293.13 & 0.00 & 0.00 & 0.68\\
\hspace{1em}2 & Skew normal & -12488.28 & -195.15 & 21.36 & 1.03\\
\hspace{1em}3 & Gamma & -12500.93 & -207.80 & 22.18 & 1.03\\
\hspace{1em}4 & Normal & -12508.91 & -215.78 & 23.29 & 1.28\\
\hspace{1em}5 & Beta & -12537.14 & -244.01 & 25.41 & 1.22\\
\addlinespace[0.3em]
\multicolumn{6}{l}{\textbf{Manicaland Female 45-49}}\\
\hspace{1em}1 & Sinh-arcsinh & -9183.03 & 0.00 & 0.00 & 0.56\\
\hspace{1em}2 & Skew normal & -9455.87 & -272.83 & 23.57 & 1.68\\
\hspace{1em}3 & Normal & -9477.33 & -294.30 & 23.55 & 1.62\\
\hspace{1em}4 & Gamma & -9497.31 & -314.27 & 25.08 & 1.44\\
\hspace{1em}5 & Beta & -9576.44 & -393.40 & 32.07 & 1.94\\
\bottomrule
\end{tabular}
\end{table}
\begin{table}[H]

\caption{\label{tab:TableManicalandMale}Full ELPD and QQ RMSE table for men in the Manicaland data set. Higher ELPD values and lower QQ RMSE values are better.}
\centering
\begin{tabular}[t]{rlrrrr}
\toprule
Rank & Distribution & ELPD & ELPD Diff & SE of Diff & QQ RMSE\\
\midrule
\addlinespace[0.3em]
\multicolumn{6}{l}{\textbf{Manicaland Male 20-24}}\\
\hspace{1em}1 & Sinh-arcsinh & -9770.00 & 0.00 & 0.00 & 0.30\\
\hspace{1em}2 & Skew normal & -9895.82 & -125.83 & 33.35 & 0.40\\
\hspace{1em}3 & Normal & -10139.11 & -369.11 & 79.13 & 0.49\\
\hspace{1em}4 & Beta & -10587.64 & -817.64 & 181.23 & 0.56\\
\hspace{1em}5 & Gamma & -11594.58 & -1824.59 & 388.26 & 1.15\\
\addlinespace[0.3em]
\multicolumn{6}{l}{\textbf{Manicaland Male 25-29}}\\
\hspace{1em}1 & Sinh-arcsinh & -13978.59 & 0.00 & 0.00 & 0.40\\
\hspace{1em}2 & Skew normal & -13990.39 & -11.80 & 8.51 & 0.48\\
\hspace{1em}3 & Normal & -14018.60 & -40.00 & 17.48 & 0.45\\
\hspace{1em}4 & Beta & -14152.35 & -173.76 & 48.77 & 0.40\\
\hspace{1em}5 & Gamma & -14500.47 & -521.87 & 117.58 & 0.55\\
\addlinespace[0.3em]
\multicolumn{6}{l}{\textbf{Manicaland Male 30-34}}\\
\hspace{1em}1 & Sinh-arcsinh & -12949.24 & 0.00 & 0.00 & 0.31\\
\hspace{1em}2 & Skew normal & -13016.44 & -67.21 & 25.01 & 0.37\\
\hspace{1em}3 & Normal & -13037.46 & -88.22 & 16.57 & 0.49\\
\hspace{1em}4 & Beta & -13070.31 & -121.07 & 54.74 & 0.41\\
\hspace{1em}5 & Gamma & -13285.47 & -336.23 & 171.92 & 0.42\\
\addlinespace[0.3em]
\multicolumn{6}{l}{\textbf{Manicaland Male 35-39}}\\
\hspace{1em}1 & Sinh-arcsinh & -11496.14 & 0.00 & 0.00 & 0.27\\
\hspace{1em}2 & Skew normal & -11528.36 & -32.22 & 9.83 & 0.39\\
\hspace{1em}3 & Normal & -11530.43 & -34.29 & 9.72 & 0.26\\
\hspace{1em}4 & Gamma & -11531.75 & -35.61 & 12.47 & 0.24\\
\hspace{1em}5 & Beta & -11582.63 & -86.49 & 12.97 & 0.48\\
\addlinespace[0.3em]
\multicolumn{6}{l}{\textbf{Manicaland Male 40-44}}\\
\hspace{1em}1 & Sinh-arcsinh & -8714.06 & 0.00 & 0.00 & 0.35\\
\hspace{1em}2 & Skew normal & -8749.78 & -35.72 & 10.11 & 0.51\\
\hspace{1em}3 & Gamma & -8777.08 & -63.02 & 10.38 & 0.55\\
\hspace{1em}4 & Normal & -8791.45 & -77.38 & 12.23 & 0.76\\
\hspace{1em}5 & Beta & -8860.22 & -146.16 & 18.15 & 0.93\\
\addlinespace[0.3em]
\multicolumn{6}{l}{\textbf{Manicaland Male 45-49}}\\
\hspace{1em}1 & Sinh-arcsinh & -7110.27 & 0.00 & 0.00 & 0.42\\
\hspace{1em}2 & Skew normal & -7177.03 & -66.75 & 25.08 & 0.76\\
\hspace{1em}3 & Gamma & -7213.99 & -103.72 & 13.02 & 1.07\\
\hspace{1em}4 & Normal & -7232.04 & -121.77 & 13.09 & 1.28\\
\hspace{1em}5 & Beta & -7312.35 & -202.08 & 18.61 & 1.61\\
\bottomrule
\end{tabular}
\end{table}

\begin{table}

\caption{\label{tab:ELPDdistTables}LOO-CV estimated ELPD values, differences, and standard errors of differences, as well as QQ RMSE values, for all five regression models fit to all four data sets. The "difference" value of a row is the difference between that row's ELPD value and data set-specific best ELPD value. Higher ELPD values and lower QQ RMSE values are better.}
\centering
\begin{tabular}[t]{rlrrrr}
\toprule
Rank & Model & ELPD & ELPD Diff & SE of Diff & QQ RMSE\\
\midrule
\addlinespace[0.3em]
\multicolumn{6}{l}{\textbf{AHRI}}\\
\hspace{1em}1 & Distributional 4 & 55841.91 & 0.00 & 0.00 & 0.66\\
\hspace{1em}2 & Distributional 3 & 55534.16 & -307.75 & 32.36 & 0.93\\
\hspace{1em}3 & Distributional 2 & 54794.79 & -1047.12 & 51.69 & 1.21\\
\hspace{1em}4 & Distributional 1 & 54335.19 & -1506.72 & 72.32 & 1.15\\
\hspace{1em}5 & Conventional & 52689.21 & -3152.70 & 100.59 & 1.30\\
\addlinespace[0.3em]
\multicolumn{6}{l}{\textbf{AHRI Deaheaped}}\\
\hspace{1em}1 & Distributional 4 & 58503.98 & 0.00 & 0.00 & 0.62\\
\hspace{1em}2 & Distributional 3 & 58219.23 & -284.75 & 28.64 & 0.92\\
\hspace{1em}3 & Distributional 2 & 57503.68 & -1000.30 & 47.14 & 1.14\\
\hspace{1em}4 & Distributional 1 & 57097.39 & -1406.59 & 64.48 & 1.06\\
\hspace{1em}5 & Conventional & 55296.25 & -3207.73 & 99.42 & 1.26\\
\addlinespace[0.3em]
\multicolumn{6}{l}{\textbf{Haiti 2016-17 DHS }}\\
\hspace{1em}1 & Distributional 4 & 5207.57 & 0.00 & 0.00 & 0.84\\
\hspace{1em}2 & Distributional 3 & 5196.69 & -10.89 & 6.54 & 0.91\\
\hspace{1em}3 & Distributional 1 & 5140.77 & -66.80 & 12.27 & 0.98\\
\hspace{1em}4 & Distributional 2 & 5138.75 & -68.83 & 12.24 & 0.99\\
\hspace{1em}5 & Conventional & 4777.78 & -429.80 & 30.54 & 1.33\\
\addlinespace[0.3em]
\multicolumn{6}{l}{\textbf{Manicaland}}\\
\hspace{1em}1 & Distributional 4 & 24516.15 & 0.00 & 0.00 & 1.04\\
\hspace{1em}2 & Distributional 3 & 24313.74 & -202.40 & 20.52 & 1.34\\
\hspace{1em}3 & Distributional 2 & 23472.07 & -1044.08 & 47.77 & 1.80\\
\hspace{1em}4 & Distributional 1 & 23192.49 & -1323.66 & 54.97 & 1.89\\
\hspace{1em}5 & Conventional & 21011.29 & -3504.86 & 89.01 & 2.05\\
\bottomrule
\end{tabular}
\end{table}

\end{document}